\documentclass[ip,twocolumn]{jpsj3}
\usepackage{txfonts}
\usepackage{amsmath}

\usepackage{graphicx,color}
\usepackage{dcolumn}
\usepackage{bm}
\usepackage{mathtools}
\usepackage{tabularx}
	\newcolumntype{C}{>{\centering\arraybackslash}X}
	\newcolumntype{L}{>{\raggedright\arraybackslash}X}
	\newcolumntype{R}{>{\raggedleft\arraybackslash}X}

\title{Neutron Scattering Study on Yttrium Iron Garnet for Spintronics}

\author{Yusuke Nambu$^1$\thanks{nambu@tohoku.ac.jp}$^{,2,3}$ and Shin-ichi Shamoto$^{4,5}$\thanks{s\_shamoto@cross.or.jp}$^{,6,7}$}
\inst{
$^1$Institute for Materials Research, Tohoku University, Sendai 980-8577, Japan \\
$^2$FOREST, Japan Science and Technology Agency, Kawaguchi, Saitama 332-0012, Japan \\
$^3$Organization for Advanced Studies, Tohoku University, Sendai 980-8577, Japan \\
$^4$Department of Physics, National Cheng Kung University, Tainan 70101, Taiwan \\
$^5$Neutron Science and Technology Center, Comprehensive Research Organization for Science and Society (CROSS), Tokai, Ibaraki 319-1106, Japan \\
$^6$Advanced Science Research Center, Japan Atomic Energy Agency (JAEA), Tokai, Ibaraki 319-1195, Japan \\
$^7$Meson Science Laboratory, RIKEN, Wako, Saitama 351-0198, Japan
}

\abst{
Spin current--a flow of the spin degree of freedom in matter--has vital importance in spintronics.
Propagation of the spin current ranges over a whole momentum space; however, generated spin currents are mainly detected in the long-wavelength limit.
To facilitate practical uses of spintronics and magnonics, microscopic understanding of the spin current is necessary.
We here address yttrium iron garnet, which is a well-employed ferrimagnet for spintronics, and review {\it in re} the momentum- and energy-resolved characteristics of its magnetism.
Using {\it unpolarized} neutrons, we refined its detailed crystal and magnetic structure, and examined magnetic excitations through four decades (10~$\mu$eV--100~meV) using chopper spectrometers in J-PARC, Japan.
We also measured mode-resolved directions of the precessional motion of the magnetic moment, i.e., magnon polarization, which carries the spin current in insulators through {\it polarized} neutron scattering, using a triple-axis spectrometer in ILL, France.
The magnon polarization is a hitherto untested fundamental property of magnets, affecting the thermodynamic properties of the spin current.
Our momentum- and energy-resolved experimental findings provide an intuitive understanding of the spin current and demonstrate the importance of neutron scattering techniques for spintronics and magnonics.
}

\begin{document}
\maketitle

\section{Introduction}

Spintronics, aiming at utilizing the spin degree of freedom instead of or in addition to the charge one of an electron, has attracted much attention recently.
Creation, annihilation, and control of the spin current--a flow of the spin degree of freedom--have been the main topics in this research field.
Spin currents can be generated electromagnetically~\cite{Silsbee1979,Mizukami2002,Tserkovnyak2002,Kato2004,Wunderlich2005}, optically,~\cite{Prins1995} and thermally~\cite{Uchida2008,Slachter2010,Cornelissen2015}, and their propagation spans the whole momentum ($\vec{Q}$) space.
However, the detection has been limited to the long-wavelength limit ($Q=0$) by voltage measurement through the inverse spin Hall effect~\cite{Azevedo2005,Saitoh2006,Valenzuela2006,Kimura2007}.
The measured voltage is the macroscopic sum of the induced spin currents; hence, only the relative intensity and overall propagation direction can be discriminated.
To gain microscopic views of the spin current including the diffusion length, lifetime, and quantitative signal strength with the effects of thermal activation, information at the characteristic momentum/energy points will be needed.
In this review, we tackle the microscopic view of the spin current from the $(\vec{Q},E)$-resolved information via neutron scattering techniques using the quintessential ferrimagnet yttrium iron garnet.

Yttrium iron garnet (YIG) with the chemical composition of Y$_3$Fe$_5$O$_{12}$ is an insulator with a complex structure.
It is an essential material for microwave and optical technologies~\cite{Wu2013} and also for basic research in spintronics, magnonics, and quantum information~\cite{Tabuchi2015}.
This is owing to the highest quality magnetization dynamics among the known magnets, yielding long magnon lifetimes~\cite{Chang2014}.
The unit cell in YIG contains Fe$^{3+}$ local moments with spin $S=5/2$ in tetrahedral ($24d$ Wyckoff position) and octahedral ($16a$ Wyckoff position) oxygen ligands with opposite spin projections in a ratio of 3:2, giving a net magnetization.
There are two major magnon dispersions, acoustic and optical modes.
The gap separating the optical and acoustic modes is on the order of the thermal energy at room temperature.
At low temperatures, YIG behaves like a simple ferromagnet with an approximately quadratic magnon dispersion.
At higher temperatures, non-parabolicities become apparent, and optical modes start to become occupied.

Neutron scattering is the unrivaled method of choice to measure the magnon dispersion across large areas of reciprocal space, enabling the magnon dispersion in magnets to be unambiguously determined.
Here, we review the crystal/magnetic structure and magnon dispersion relations of YIG in a wide $(\vec{Q},E)$-regime and the mode-resolved direction of the precessional motion of the magnetic moments, i.e., magnon polarization.
Results are mainly based on both unpolarized~\cite{Shamoto2018,Shamoto2020} and polarized~\cite{Nambu2020} neutron scattering experiments.
We find negatively polarized optical modes over the exchange gap, as well as the positively polarized acoustic mode, confirming the ferrimagnetic character in YIG, and that thermal excitation of the optical mode will limit the amplitude of the induced spin current~\cite{Nambu2020}.

This review is structured as follows.
We first discuss the detailed crystal and magnetic structure of YIG.
We then show magnetic excitations in YIG ranging from high (100~meV) to ultralow (10~$\mu$eV) energy with macroscopic magnetization and specific heat results.
Next, we move on to polarized neutron scattering results, starting with a review of the magnon polarization as the spin current carrier.
Cross sections as a function of the direction of neutron polarization are then explained, which are needed to follow the results of the polarized neutron scattering.
Finally, the mode-resolved magnon polarization is explicitly determined through the chiral term detection.
The authors wrote Section 5 together and separately wrote other Sections: Y.~N. is responsible for Sects. 1, 4, and 6, and S.~S. is responsible for Sects. 2 and 3.

\section{Crystal and Magnetic Structure}

\begin{figure}[t]
	\begin{center}
		\includegraphics[width=0.8\linewidth,clip]{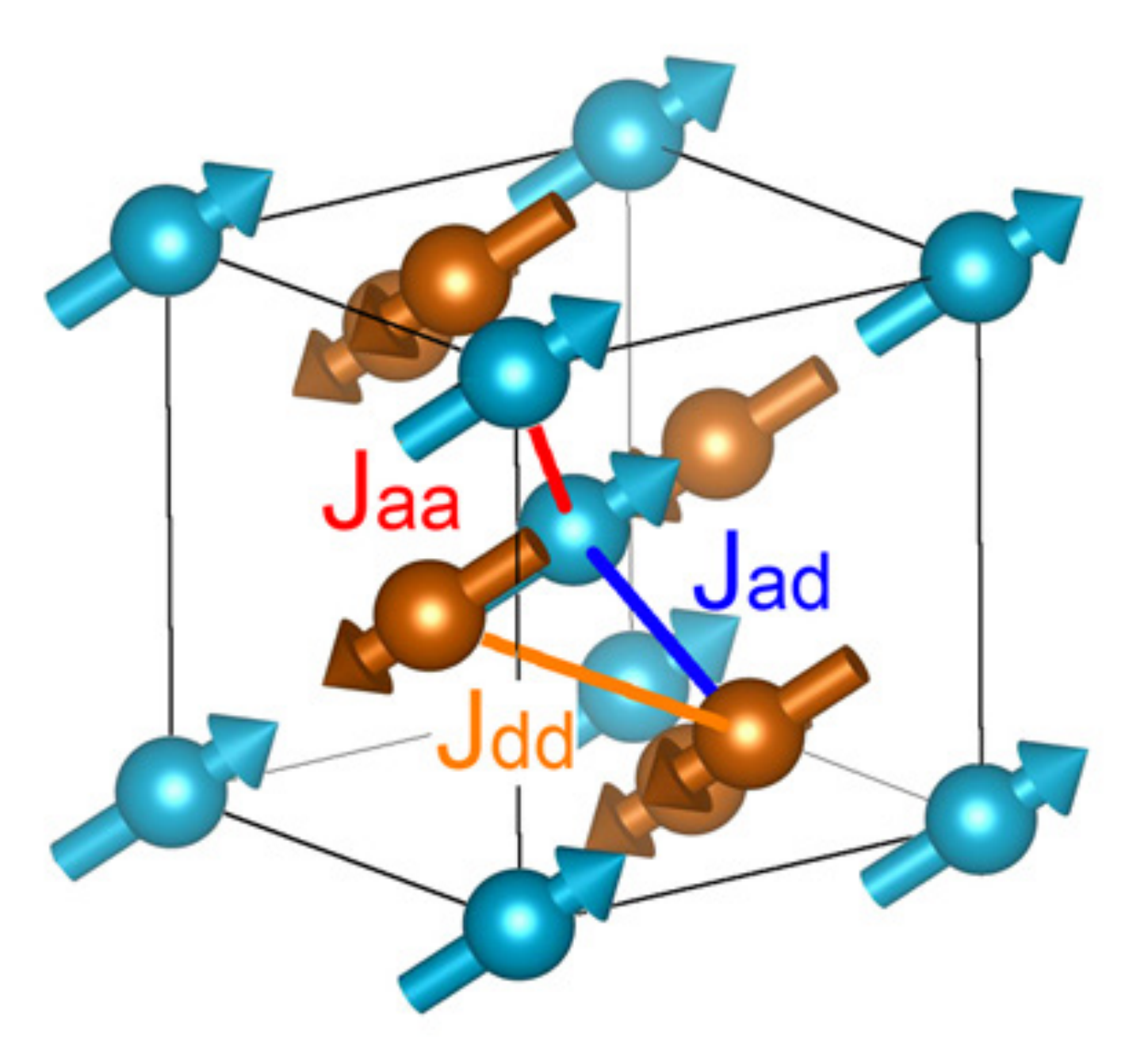}
	\end{center}
	\caption{(Color online) Fe spins in a sub-unit cell that corresponds to $1/8$ of a cubic unit cell ($Ia\bar{3}d$) with 40 Fe spins. Blue and brown arrows are spins at 16$a$ (octahedral) and 24$d$ (tetrahedral) sites for $Ia\bar{3}d$, respectively. Three nearest-neighbor-exchange integrals between 16$a$ and 24$d$ sites, $J_{aa}$, $J_{ad}$, and $J_{dd}$, are shown by red, blue, and orange lines, respectively. Reprinted with permission from Shamoto {\it et al}.~\cite{Shamoto2018}({\copyright}$\,$ 2018 The American Physical Society).}
	\label{fig:1}
\end{figure}

Recent detailed studies of the longitudinal spin Seebeck effect (LSSE) on YIG have revealed the importance of the basic parameters of YIG under magnetic fields~\cite{Kikkawa2015,Kikkawa2016}. 
Here, the crystal and magnetic structure is discussed in detail.  
In a sub-unit cell of this ferrimagnet with five Fe spins, as shown in Fig.~\ref{fig:1}, there are three up spins and two down spins, corresponding to the three positive and two negative polarization modes, respectively.
A large number of basic properties of YIG have been historically reported~\cite{Cherepanov}.
The crystal and magnetic structures were studied by using a powder sample~\cite{Rodic}.
The crystal structure of YIG was distorted from cubic to trigonal symmetries under a magnetic field of 0.2~T~\cite{Rodic}.
For precise crystal and magnetic structure refinements, a magnetic field is required to remove the magnetic domain walls from a single crystal. 
Such a measurement was carried out at about 295~K under a magnetic field ($B\approx 0.1$~T) along [111]$_{\rm cubic}$ with a pair of permanent magnets.
The magnetic field at the sample position was measured using a Hall effect sensor.
Single crystals were grown by a traveling solvent floating zone method~\cite{Kimura} with an image furnace with four halogen lamps (FZ-T-4000-H-II-S-TS, Crystal Systems Co., Ltd.).
This method allows us to grow impurity-free crystals. Under a magnetic field, the demagnetization effect can become large for a ferromagnet, depending on the shape. In our neutron scattering experiments, each rod crystal was grown along each magnetic field to reduce the demagnetization effect.

\begin{figure}[t]
	\begin{center}
		\includegraphics[width=0.8\linewidth,clip]{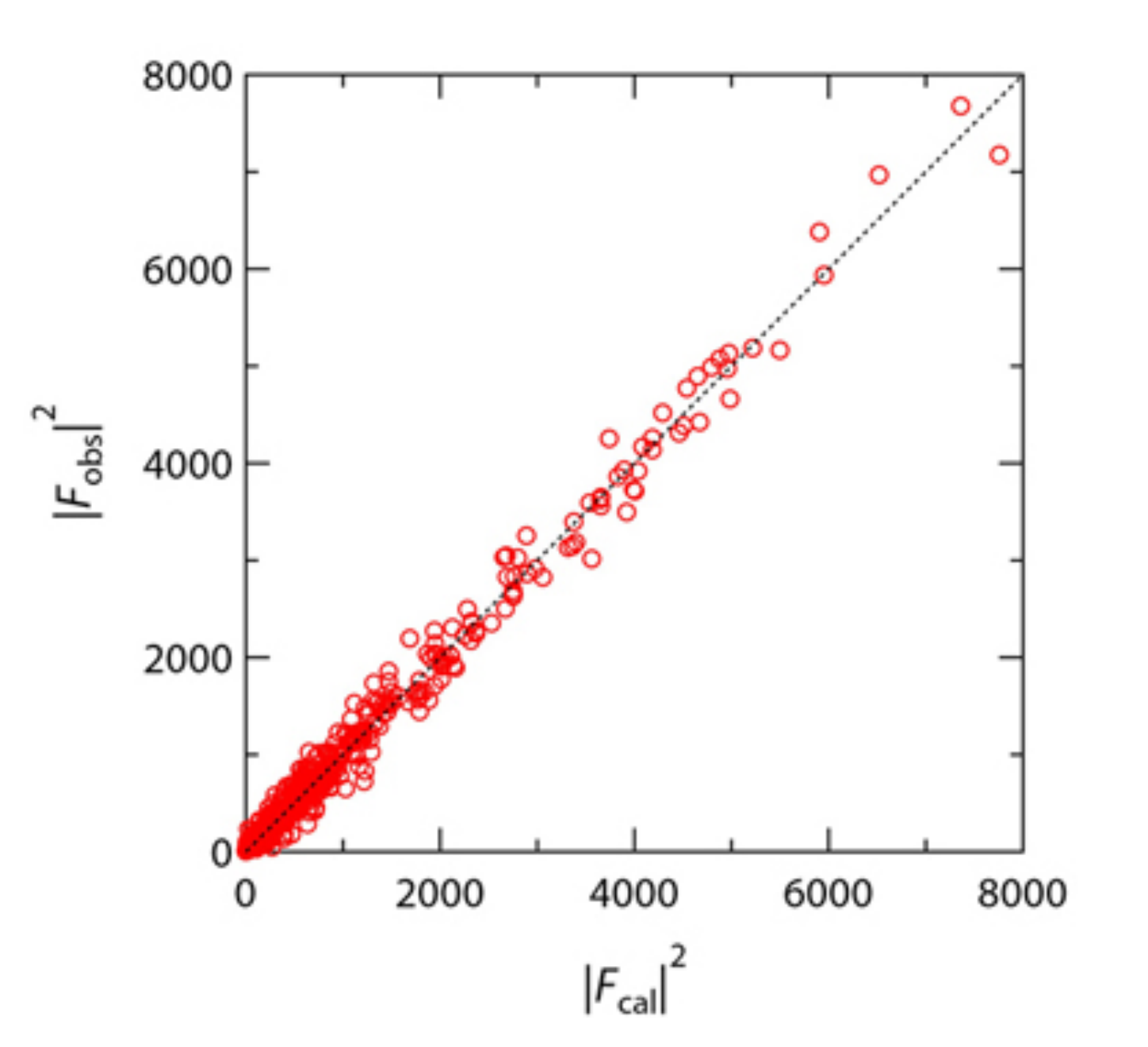}
	\end{center}
	\caption{(Color online) Observed nuclear and magnetic Bragg peak intensity refined as a function of calculated intensity refined with the trigonal space group ($R\bar{3}$). Reprinted with permission from Shamoto {\it et al}.~\cite{Shamoto2018} ({\copyright}$\,$ 2018 The American Physical Society).}
	\label{fig:2}
\end{figure}

Nuclear and magnetic Bragg reflections were measured at BL18 SENJU~\cite{SENJU} of J-PARC MLF.
Their intensities were refined with a trigonal space group ($R\bar{3}$, No.~148: hexagonal setting)~\cite{Rodic} with lattice parameters $a=17.50227(55)$~\AA \ and $c=10.73395(29)$~\AA$\;$ by using {\sc FullProf Suite}~\cite{FullProf} and {\sc STARGazer} software~\cite{STARGazer}.
The structural refinement result is shown in Fig.~\ref{fig:2}.

\begin{figure}[b]
	\begin{center}
		\includegraphics[width=0.8\linewidth,clip]{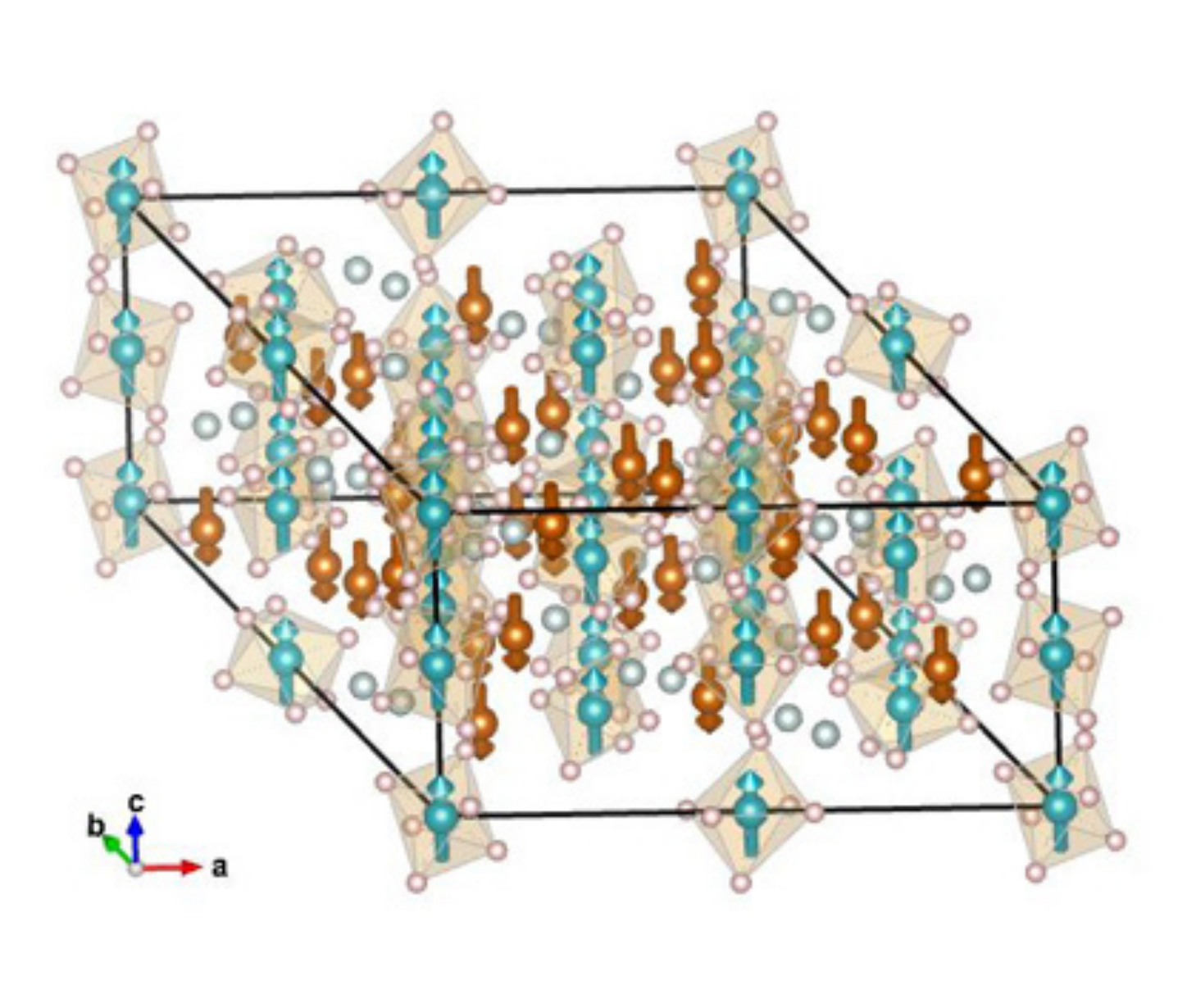}
	\end{center}
	\caption{(Color online) Refined crystal and magnetic structure of YIG in the trigonal unit cell of $R\bar{3}$. Blue and brown arrows are iron spins for the octahedral and tetrahedral sites, respectively, small pink spheres are oxygen, and pale blue spheres are yttrium. The structure was drawn using {\sc VESTA} software~\cite{VESTA}. Reprinted with permission from Shamoto {\it et al}.~\cite{Shamoto2018} ({\copyright}$\,$ 2018 The American Physical Society).}
	\label{fig:3}
\end{figure}

All the magnetic moments align along [111]$_{\rm cubic}$ ([001]$_{\rm hexagonal}$) parallel to the applied magnetic field $B$.
The refined crystal and magnetic structure is shown in Fig.~\ref{fig:3}.
The refined crystallographic parameters with reliability factors $R_{F^2}=9.85$\% and $R_F=7.07$\% are listed in Table~\ref{table1}.
\begin{table*}[h]
	\caption{Parameters of the crystal and magnetic structure of YIG at about 295~K under $B\approx 0.1$~T in the space group $R\bar{3}$. Errors are shown in parentheses by the corresponding digits. Note that the occupancies $g$ of Fe and O7$_{18f}$ (indicated by ``fix'') were fixed during the refinements.} 
	\label{table1}
	\centering
	\begin{tabularx}{\textwidth}{lRRRRRR}
		\hline\hline
		Atoms	& \multicolumn{3}{c}{Fractional coordinates}	& $B$~(\AA$^2$)	& $g$	& $\mu_z$~(unit of ${\rm \mu_B}$)	\\
				& \multicolumn{1}{c}{$x$}	& \multicolumn{1}{c}{$y$}	& \multicolumn{1}{c}{$z$}	&	&	&	\\
		\hline
		Y1$_{18f}$	& 0.1255(3)	& 0.0005(3)	& 0.2497(2)	& 0.245(53)	& 0.943(22)	&	\\
		Y2$_{18f}$	& 0.2911(3)	& 0.3333(3)	& 0.5834(2)	& 0.245(53)	& 0.948(22)	&	\\
		Fe$_{3a}$	& 0	& 0	& 0	& 0.243(27)	& 1.00(fix)	& 3.50(17)	\\
		Fe$_{3b}$	& 0	& 0	& 0.5	& 0.243(27)	& 1.00(fix)	& 3.50(17)	\\
		Fe$_{9d}$	& 0	& 0.5	& 0.5	& 0.243(27)	& 1.00(fix)	& 3.50(17)	\\
		Fe$_{9e}$	& 0.5	& 0	& 0	& 0.243(27)	& 1.00(fix)	& 3.50(17)	\\
		Fe1$_{18f}$	& 0.2084(2)	& 0.1672(2)	& 0.4166(2)	& 0.315(28)	& 1.00(fix)	& $-$3.37(17)	\\
		Fe2$_{18f}$	& 0.2912(2)	& $-$0.1670(2)	& 0.5832(2)	& 0.315(28)	& 1.00(fix)	& $-$3.37(17)	\\
		O1$_{18f}$	& 0.0877(3)	& 0.0920(4)	& 0.1210(2)	& 0.344(28)	& 0.993(18)	&	\\
		O2$_{18f}$	& 0.2622(4)	& 0.1158(4)	& 0.3230(3)	& 0.344(28)	& 0.941(17)	&	\\
		O3$_{18f}$	& $-$0.4212(3)	& $-$0.3721(4)	& 0.5444(2)	& 0.344(28)	& 0.991(20)	&	\\
		O4$_{18f}$	& 0.4867(4)	& 0.0953(4)	& 0.4188(3)	& 0.344(28)	& 0.940(17)	&	\\
		O5$_{18f}$	& $-$0.0042(4)	& $-$0.0904(4)	& 0.3798(3)	& 0.344(28)	& 0.963(18)	&	\\
		O6$_{18f}$	& 0.1453(4)	& $-$0.1154(4)	& 0.1773(3)	& 0.344(28)	& 0.940(18)	&	\\
		O7$_{18f}$	& $-$0.0490(3)	& 0.3717(4)	& $-$0.0451(2)	& 0.344(28)	& 1.00(fix)	&	\\
		O8$_{18f}$	& 0.3899(4)	& $-$0.0967(4)	& 0.0809(3)	& 0.344(28)	& 0.933(17)	&	\\
		\hline\hline
	\end{tabularx}
\end{table*}
They were almost consistent with the reported parameters~\cite{Rodic}.
The occupancy $g$ of O7$_{18f}$ was fixed because of the slightly larger $g$ value than 1.00 within the error during the refinement.
The chemical composition of the present YIG crystal was Y$_{2.84(9)}$Fe$_{5}$O$_{11.57(21)}$.
The deficiency of the Y$^{3+}$ ion was almost compensated by the oxygen deficiency for the Fe valence of $+3$ within the error.
The obtained magnetic moments, with ${\rm \mu_B}$ being the Bohr magneton, were $3.50\pm 0.17$ ${\rm \mu_B}$ and $3.37\pm0.17$ ${\rm \mu_B}$ at the octahedral and tetrahedral sites, respectively.
Although the obtained magnetic moments under $B\approx 0.1$~T are smaller than $4.47\pm 0.04$~${\rm \mu_B}$ and $4.02\pm 0.05$~${\rm \mu_B}$~\cite{Rodic}, the total magnetization of 3.1(6) ${\rm \mu_B}$/f.u. agrees with the magnetization of 3.05~${\rm \mu_B}$/f.u. under $B = 1$~T.
In the previous study~\cite{Rodic}, magnetic domain walls may have remained in the powder sample, resulting in these discrepancies.
The slightly larger trigonal lattice distortions observed here compared with the previous ones reduce the observed magnetic moments in the present analysis due to the overlapping of Bragg peaks.
This result suggests that it is important to determine the crystal structure of YIG together with the estimation of the magnetic moments under a magnetic field.
Regardless of the atomic distortions from cubic to trigonal symmetries, the following magnon dispersions are discussed in the cubic symmetry of $Ia\bar{3}d$ (No.~230) for simplicity.   

\section{Inelastic $\bm{Unpolarized}$ Neutron Scattering}
\subsection{High-energy spin-wave excitations}

According to detailed studies of LSSE on YIG, LSSE depends on the detailed structure of the magnon density of states (MDOS, ${\mathcal D}_M$)~\cite{Kikkawa2015,Kikkawa2016}.
The magnetic field produces a Zeeman energy gap in the ferromagnetic dispersion, resulting in the reduction of thermally excited spin current.
The theoretical model is based on a simple quadratic magnon dispersion of YIG measured by previous inelastic neutron scattering (INS) measurements~\cite{Plant1977,Plant2}.
The magnon dispersion relations of YIG were first studied up to 55~meV by INS measurements~\cite{Plant2}.
The exchange integrals of YIG have been estimated under the energy limitation~\cite{Plant1977,Plant2,Cherepanov,Serga}.
Owing to high-efficiency INS spectrometers at pulsed-neutron sources, it has become possible to access high energies above 55~meV even with a small crystal.
A recent detailed magnon study was still limited to about 80~meV~\cite{Princep2017a}.
A measurement covering the full range of the energy of interest of YIG was first carried out by Shamoto's group~\cite{Shamoto2018}.
The theory of YIG magnons for LSSE~\cite{Barker} emphasizes the importance of the mode mixing for the lowest-$E$ branches.
For an antiferromagnet, the lowest-$E$ dispersion is known to have doubly degenerate modes.
For this ferrimagnet YIG, however, the lowest-$E$ dispersion is theoretically predicted to have only a single mode of positive polarization~\cite{Barker}.
The mode number can be verified from the MDOS (${\mathcal D}_M$) if we measure it on an absolute scale. Because of the estimation difficulties, there has been no report of the absolute MDOS for YIG so far.
For the MDOS estimation, we introduce an approximated dynamical structure factor and effective reciprocal space volume to simplify the absolute estimation. In addition, we performed the numerical calculation of absorption coefficients.
The INS probability on a magnet can be expressed by Fermi's golden rule, including the MDOS of the final states.
This is the same as the phonon density of states for the phonon dispersions, which is often measured by INS.
By using this method, the MDOS of YIG was estimated from the observed scattering intensity.
For the inelastic magnetic scattering, the sum of the $q$-integrated scattering function $S(E)$ after energy integration is well known to be proportional to $S(S+1)$~\cite{Shamoto2018}.
Therefore, the $q$-integrated dynamical spin susceptibility $\chi^{\prime\prime}(E)$ normalized by $g^2{\rm \mu^2_B}S(S+1)$ gives the MDOS at $T=0$~K, where the energy integration becomes unity. 
The simple quadratic dispersion model~\cite{Kikkawa2015} at low energies below 14~meV has been successfully used to estimate the MDOS, suggesting the validity of our estimation.
The observed lowest-$E$ dispersion was fitted by a simple quadratic dispersion with a stiffness constant $D$.
The fitted parameter $D$ can also be verified by the exchange integrals obtained from the whole magnon spectrum in our experiment.
Thus, the relation between the MDOS and $\chi^{\prime\prime}(E)$ is discussed quantitatively on the basis of the stiffness constant $D$. 

Here, the Heisenberg Hamiltonian of the spin system in YIG is written as~\cite{Cherepanov}
\begin{equation}
\mathcal{H}=-2\sum_{ij} J_{ij}S_{i}S_{j} + g\mu_{B} B\sum_{i}S_{i} + \sum_{i}KS_{zi}^2, 
\label{eq:1}
\end{equation}
where $S=5/2$, $g=2$ for the Fe$^{3+}$ spins, $B$ is the magnetic field, and $K$ is the anisotropy coefficient.
\begin{figure*}[t]
	\begin{center}
		\includegraphics[width=0.8\linewidth,clip]{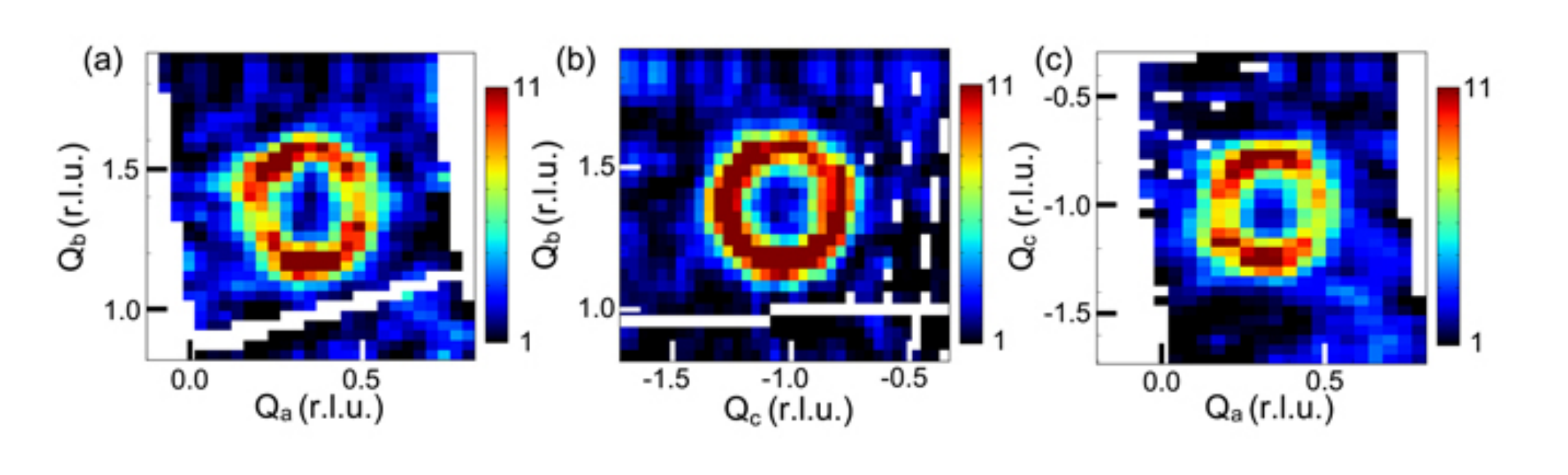}
	\end{center}
	\caption{(Color online) One of the magnon excitations of YIG at (220) ($(Q_a, Q_b, Q_c)$=(1/3, 4/3, $-$1)) in the $E$ range from 5 to 20~meV measured with $E_{\rm i}=45.3$~meV. (a) $Q_a$--$Q_b$ contour map of $-1.05<Q_c<-0.95$. (b) $Q_c$--$Q_b$ contour map of $0.3<Q_a<0.4$. (c) $Q_a$--$Q_c$ contour map of $1.3<Q_b<1.4$. White areas are regions not covered by the detector. The color bars are in the unit of mbarn sr$^{-1}$meV$^{-1}$r.l.u.$^{-3}$. Reprinted with permission from Shamoto {\it et al}.\cite{Shamoto2018} ({\copyright}$\,$ 2018 The American Physical Society).}
	\label{fig:4}
\end{figure*}
Although complex models with a large number of parameters have been intensively studied~\cite{Plant2,Princep2017a}, a minimum model has three nearest-neighbor exchange integrals, $J_{ad}$, $J_{aa}$, and $J_{dd}$, where the subscripts $a$ and $d$ refer to the Fe 16$a$ (octahedral) and 24$d$ (tetrahedral) sites in the cubic symmetry  $Ia\bar{3}d$, respectively.
The spin Hamiltonian was diagonalized using the {\sc spinW} software package~\cite{spinw} based on the linear spin-wave theory with the Holstein--Primakoff approximation.
The spin-wave spectra were drawn by using the {\sc Horace} software package~\cite{Horace}.

A magnon excitation in YIG is observed at ${\bf q}$ in a reciprocal space deviating from the $\Gamma$ point at a finite energy transfer $E$ by using the BL01 4SEASONS~\cite{Kajimoto}, BL14 AMATERAS\cite{Nakajima} with the multi-$E_{\rm i}$ option~\cite{Nakamura}, and BL02 DNA~\cite{Shibata} spectrometers at J-PARC MLF.
The data sets were analyzed by using the {\sc Utsusemi} software package~\cite{Utsusemi}.
It forms a three-dimensional (3D) spherical shell at $E$ in the reciprocal ${\bf Q}$-space due to the nearly isotropic 3D interactions of localized spins as shown in Fig.~\ref{fig:4}.
Here, we define the scattering wave vector ${\bf Q}$ as ${\bf Q}={\bf q}+{\bf G}$, where ${\bf Q}=Q_a(2,-1,-1)+Q_b(1,1,1)+Q_c(0,-1,1)$, ${\bf q}=q_a(2,-1,-1)+q_b(1,1,1)+q_c(0,-1,1)$ in the crystal-setting as below Brillouin zone, and ${\bf G}$ is a reciprocal lattice vector such as (2, 2, 0).
$Q_a$, $Q_b$, $Q_c$, $q_a$, $q_b$, and $q_c$ are in reciprocal lattice units (r.l.u.).
The crystal was aligned in the ($Q_a$, $Q_b$, 0) zone as a scattering plane.
This crystal-setting Brillouin zone with 1 r.l.u.$^3$ is six times larger than the original Brillouin zone.
The $q$-integrated magnon intensity in the Brillouin zone was obtained through the integration of one 3D spherical-shell excitation.
Note that phonons were not observed in the low-$Q$ region.

\begin{figure}[b]
	\begin{center}
		\includegraphics[width=0.8\linewidth,clip]{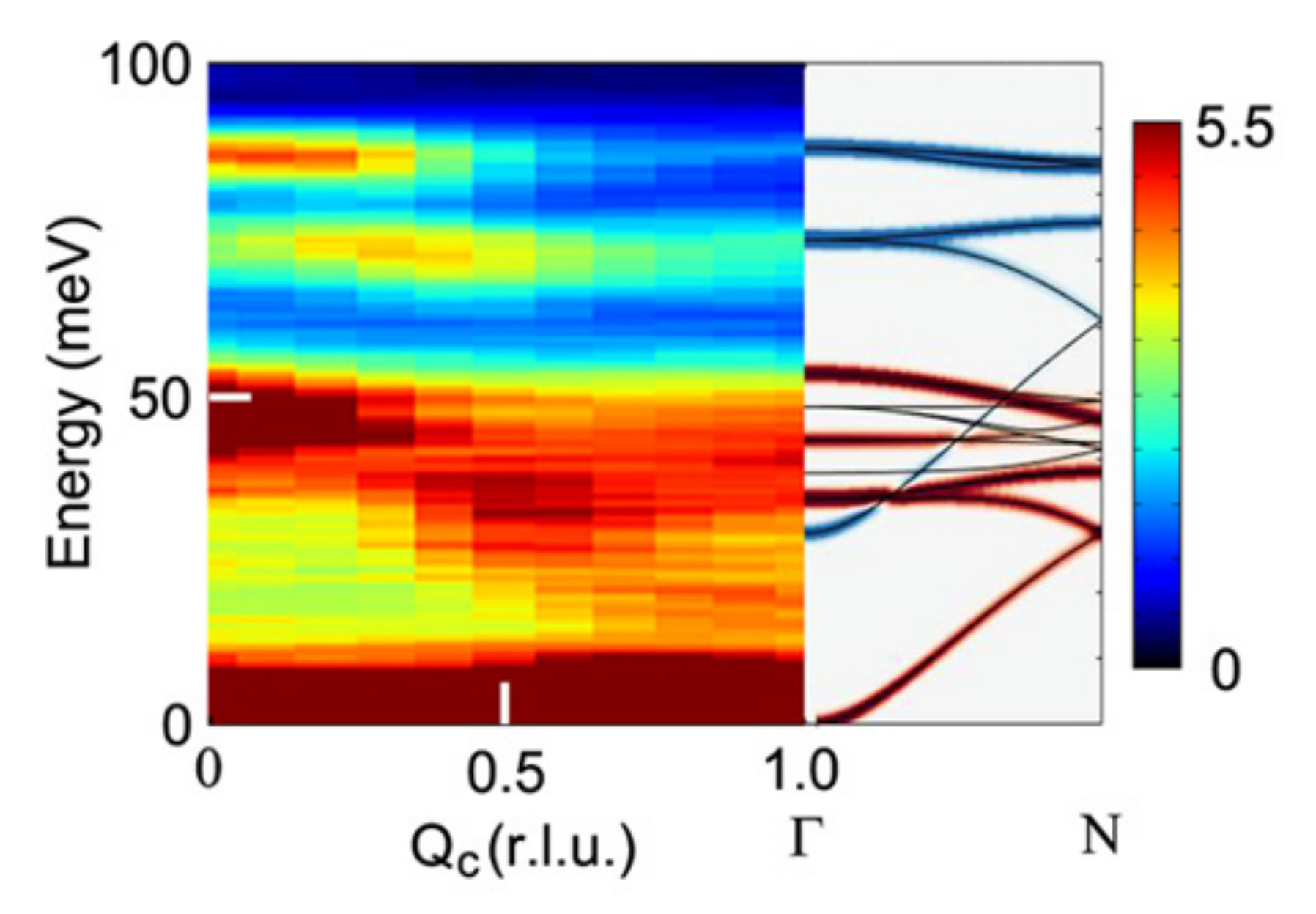}
	\end{center}
	\caption{(Color online) Wide-$E$-range magnon dispersions in the $Q_c$--$E$ space. Left: observed pattern as a function of $Q_c$ measured with $E_{\rm i}=150.0$~meV in the range of $-0.5< Q_a<3$ and $1< Q_b< 3$. Right: magnon dispersion relations along the same direction calculated from the $\Gamma$ to $N$ point at (123) by {\sc spinW} with the three nearest-neighbor exchange integrals estimated here. The brown (blue) coloring denotes the positive (negative) polarization mode. The color bar is in the unit of mbarn sr$^{-1}$meV$^{-1}$r.l.u.$^{-3}$. Reprinted with permission from Shamoto {\it et al}.~\cite{Shamoto2018} ({\copyright}$\,$ 2018 The American Physical Society).}
	\label{fig:5} 
\end{figure}
The magnons of YIG extended up to 86~meV as shown in Fig. \ref{fig:5}. The strong magnetic excitations at about 73 and 86~meV were nearly $Q$-independent in the dispersions. In the middle-$E$ range below 55~meV, the dispersions become a broad band down to 30~meV due to the overlapping of many dispersions.
Three nearest-neighbor exchange integrals, $J_{aa}$, $J_{ad}$, and $J_{dd}$, were estimated step by step by comparing with the simulation with $gS$=5 $\mu_{\rm B}$ using {\sc spinW} software~\cite{spinw} as follows. 

\begin{figure}[t]
	\begin{center}
		\includegraphics[width=0.8\linewidth,clip]{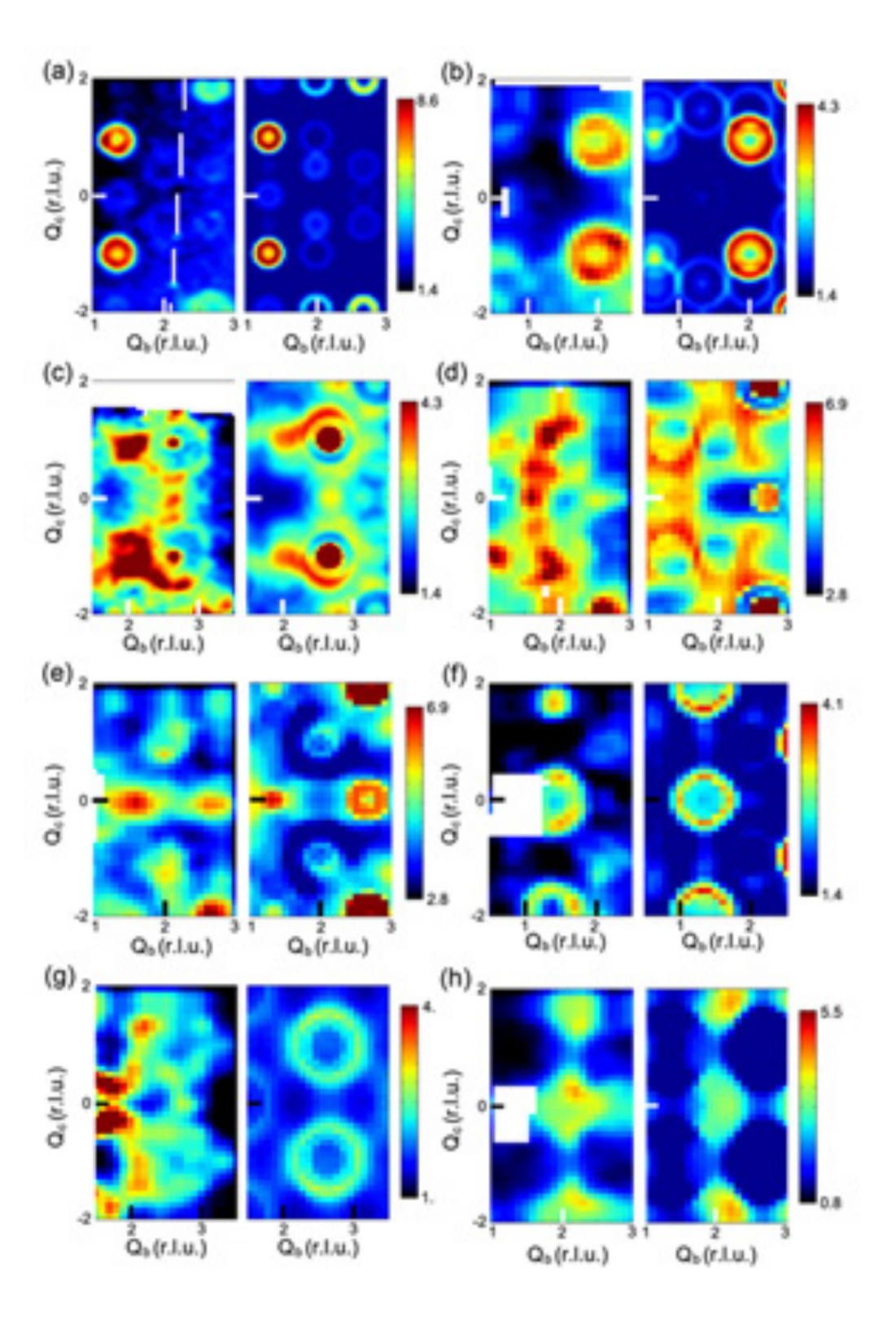}
	\end{center}
	\caption{(Color online) Constant-$E$ cuts of magnon spectra in the $Q_b$--$Q_c$ plane with 10~meV $E$ intervals. Left and right plots show observed and simulated patterns, respectively. The transfer energies are (a) 12, (b) 25, and (c) 33~meV for $E_{\rm i}=45.3$~meV and (d) 40, (e) 50, (f) 60, (g) 70, and (h) 85~meV for $E_{\rm i}=150.0$~meV. The corresponding $h$ values in (2$h$, $-h$, $-h$) are about 0.2, 0.9, 1.7, 0.65, 0.95, 1.25, 1.6, and 1.9, respectively. The color bars are for observed spectra in unit of mbarn sr$^{-1}$meV$^{-1}$r.l.u.$^{-3}$. Reprinted with permission from Shamoto {\it et al}.~\cite{Shamoto2018} ({\copyright}$\,$ 2018 The American Physical Society).}
	\label{fig:6} 
\end{figure}

$J_{ad}$ was determined from the whole magnon bandwidth.
A strong positive correlation was found between $J_{ad}$ and $J_{aa}$, which was sensitive to the second-highest magnon energy at $P$ ($\sim$70~meV). $J_{dd}$ was determined by the magnon energy at $P$ ($\sim$45~meV) with positive polarization in the middle-$E$ range.
The three nearest-neighbor exchange integrals, $J_{aa}$, $J_{ad}$, and $J_{dd}$, became $0.00\pm 0.05$, $-2.90\pm 0.07$, and $-0.35\pm 0.08$~meV, respectively.
The minus sign means that the couplings are antiferromagnetic in the definition of Eq.~\ref{eq:1}.
The errors of integrals correspond to the largest energy shift of up to 2~meV in the dispersion energies, typically at the $P$ point.
The calculated dispersion relations are shown in Fig.~\ref{fig:5}.
The present three exchange integrals agree with reported values ($J_{aa}$$\sim 0$, $J_{ad}=-2.78$, $J_{dd}=-0.28$~meV) estimated from the magnetic susceptibility above 750~K~\cite{Wojtowicz} with the temperature dependence correction of the lattice constant.
$J_{aa}$ is estimated to be less than $-0.03$~meV based on the study of the garnet compound Ca$_{3}$Fe$_{2}$Si$_{3}$O$_{12}$ with only 16$a$ Fe$^{3+}$ sites~\cite{Wojtowicz,Geller}.
In the previous measurement below 55~meV~\cite{Plant1977}, $J_{aa}$, $J_{ad}$, and $J_{dd}$ were $-0.69$, $-3.43$, and $-0.69$~meV, respectively.
After detailed refinement of the same magnon data, $J_{aa}$, $J_{ad}$, and $J_{dd}$ became $-0.33$, $-3.43$, and $-1.16$~meV, respectively~\cite{Cherepanov}.
The magnon dispersion relations simulated with these integrals seem consistent with those below 55~meV, but deviate largely from the present observed dispersions above 55~meV.
To verify the validity of our exchange integrals, observed and simulated constant-$E$ cuts at various energies are compared in Fig.~\ref{fig:6}.
They are consistent with each other even in the middle-$E$ range from 30 to 50~meV.
Precise fittings with more parameters than ours were reported in recent studies~\cite{Xie, Princep2017a}.
According to the model~\cite{Princep2017a}, however, simulated spin-waves with their exchange parameters have energies exceeding 100~meV.
On the other hand, the highest energy was 86~meV in our case.
This discrepancy may be due to the limited $E_{\rm i}$ of 120~meV at the MAPS time-of-flight neutron spectrometer at the ISIS spallation neutron source, which may not fully cover the high-energy spin-wave of YIG. 
Although our INS data with $E_{\rm i}=150$~meV covered energies of above 100~meV, no dispersion was observed above 86~meV as shown in Fig.~\ref{fig:5}. 

\begin{figure}[t]
	\begin{center}
		\includegraphics[width=0.8\linewidth,clip]{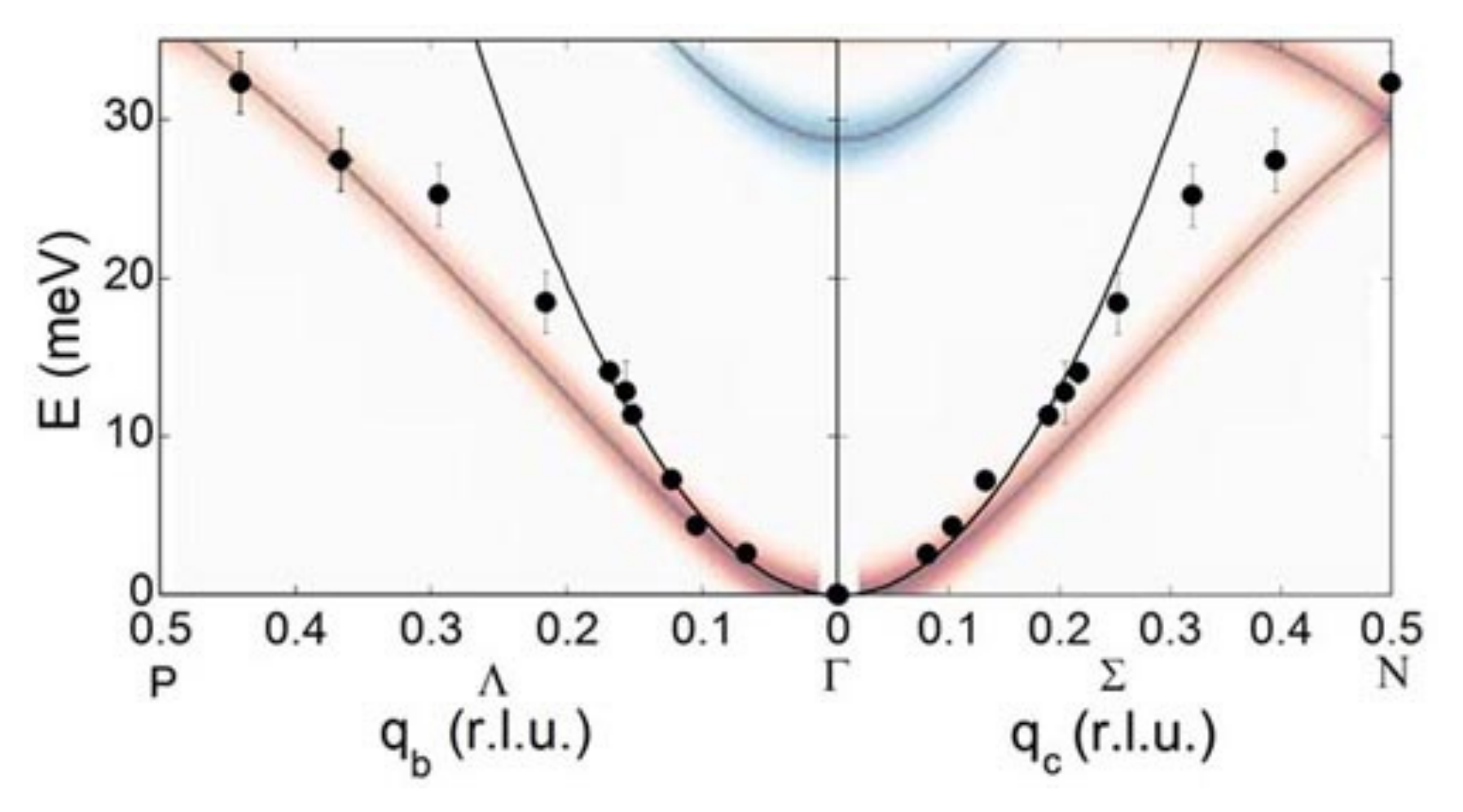}
	\end{center}
	\caption{(Color online) Lowest-$E$ magnon dispersion along the $\Lambda$ and $\Sigma$ directions measured at $\sim$20~K. The solid line is the fitting of the quadratic function of Eq.~\ref{eq:2}. The calculated dispersions with exchange integrals are also shown by pale blurry lines in the same $Q$--$E$ space. Brown (blue) denotes the positive (negative) polarization mode. Reprinted with permission from Shamoto {\it et al}.~\cite{Shamoto2018} ({\copyright}$\,$ 2018 The American Physical Society).}
	\label{fig:7}
\end{figure}

A quadratic dispersion has been observed in the lowest-$E$ acoustic magnon dispersion measured at various spectrometers below 14~meV near the $\Gamma$ point as shown in Fig.~\ref{fig:7}.
The nearly isotropic low-$E$ dispersion can be written approximately as follows:
\begin{equation}
E=Da^2q^2+E_g,
\label{eq:2}
\end{equation}
where $D$ is the stiffness constant, $q$ is the magnon wave vector, $a$ is the lattice constant, and $E_g$ is the energy gap.
The energy gap $E_g$ in YIG is approximated as the summation of the Zeeman energy by the applied magnetic field $B$ ($=\mu_{0}H$) and the anisotropy field $KS_z^2$ as follows:
\begin{equation}
E_g=g\mu_{\rm B}B+KS_z^2.
\label{eq:3}
\end{equation}
$Da^2$ is estimated to be $633\pm 17$ meV\AA$^2$ (3.95 $\times$ 10$^{-29}$ erg cm$^2$ = 3.95 $\times$ 10$^{-40}$ J m$^2$) according to the fitting below 14~meV in Fig.~\ref{fig:7}.
It is slightly smaller than the value of 670~meV\AA$^2$ (4.2 $\times$ 10$^{-29}$ erg cm$^2$) used in an LSSE study~\cite{Kikkawa2015}.
The present value is also roughly consistent with the other reported values of $580\pm 60$ meV\AA$^2$ ~\cite{Man2017} and $\sim$533 meV\AA$^2$~\cite{Cherepanov1993}.
$D$ can be estimated using the obtained exchange integrals as follows~\cite{Srivastava,Cherepanov}:
\begin{equation}
D=\frac{5}{16}\left(8J_{aa}-5J_{ad}+3J_{dd}\right).
\label{eq:4}
\end{equation}
From this equation, $Da^2$ is estimated as 642 meV\AA$^2$ from our three exchange integrals.
This value agrees with the stiffness constant obtained from Eq.~\ref{eq:2} and Fig.~\ref{fig:7}. 

\subsection{Magnon density of states}

The imaginary part of the dynamical spin susceptibility $\chi^{\prime\prime}({\bf q}, E)$ is estimated from the following equation of the magnetic differential scattering cross section:
\begin{eqnarray}
\left(\frac{{\rm d}^2\sigma}{{\rm d}\Omega{\rm d}E}\right)_{M}&=& \frac{(\gamma r_{\rm e})^2}{\pi g^2 \mu_{\rm B}^2} \frac{{\bf k}_{\rm f}}{{\bf k}_{\rm i}} f^2(Q) t^2({\bf Q}) \left\{1+(\hat{\tau} \cdot \hat{\eta})^2\right\}_{\rm av} \nonumber\\
&&\times \left\{1+n(E)\right\} \chi^{\prime\prime}({\bf Q},E)\;,
\label{eq:5}
\end{eqnarray}
where the constant value  $(\gamma r_e)^2$ is 0.2905~barn sr$^{-1}$; $g$ is the Land\'{e} $g$ factor; ${\bf k}_{\rm i}$ and ${\bf k}_{\rm f}$ are the incident and final wavenumbers, respectively; the isotropic magnetic form factor $f^2(Q)$ of Fe$^{3+}$ at (220) is 0.8059 ($Q=1.44$ \AA$^{-1}$); the dynamic structure factor $t^2({\bf Q})$~\cite{Shirane} is approximated to be the square static magnetic structure factor relative to the full moments, i.e., $t^2({\bf Q})$$\approx$$F^2_M(${\bf G}$)$/$F^2_{M0}$ =13/25; $\hat{\tau}$ is the unit vector in the direction of ${\bf Q}$; $\hat{\eta}$ is the unit vector in the mean direction of the spins; the angle-dependent term $\{1+(\hat{\tau} \cdot \hat{\eta})^2\}_{av}$ is 4/3 due to the domain average without a magnetic field; and $n(E)$ is the Bose factor.

\begin{figure}[t]
	\begin{center}
		\includegraphics[width=0.8\linewidth,clip]{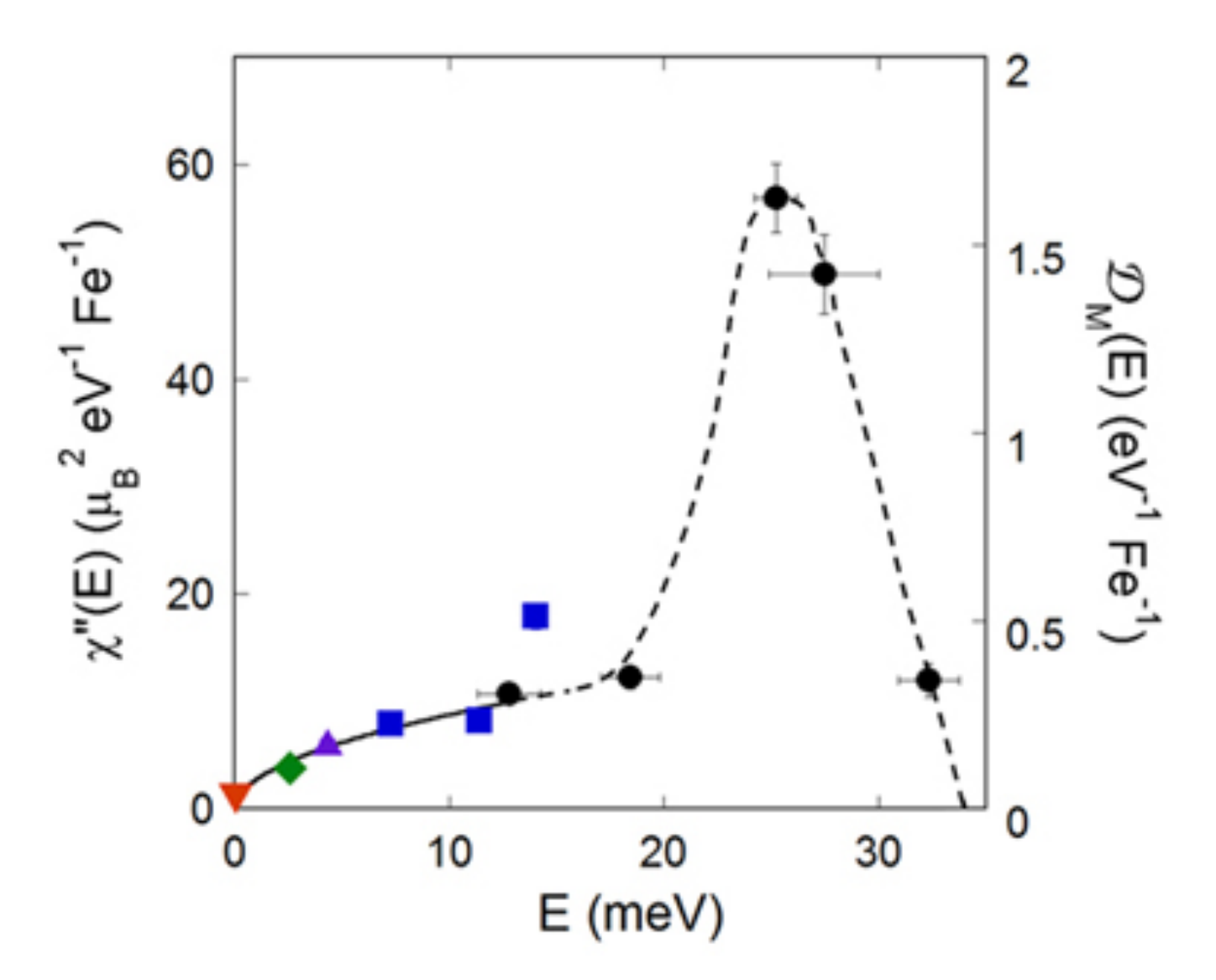}
	\end{center}
	\caption{(Color online) Energy dependence of $q$-integrated dynamical spin susceptibility $\chi^{\prime\prime}(E)$ for the lowest-$E$ magnon mode. Downward triangle: result obtained at BL02 DNA. Diamond: result obtained at BL14 AMATERAS. Upward triangle, squares, and circles: results obtained at BL01 4SEASONS from $E_{\rm i}=$ 12.5, 21.5, and 45.3~meV, respectively. The solid line is the fitting with a single parameter $\chi^{\prime\prime}_0$ using Eq.~\ref{eq:6} at $E_g$= 0. The dashed line is a guide to the eye. Reprinted with permission from Shamoto {\it et al}.~\cite{Shamoto2018}  ({\copyright}$\,$ 2018 The American Physical Society).}
	\label{fig:8}
\end{figure}

The imaginary part of the $q$-integrated dynamical spin susceptibility $\chi^{\prime\prime}(E)$ was obtained, as shown in Fig.~\ref{fig:8}.
The $E$-dependence of $\chi^{\prime\prime}(E)$ for a quadratic dispersion case becomes a square-root function of energy~\cite{Shirane}.
In the case of ferromagnetic Fe, the constant-$E$ scan intensity of magnons with a certain $E$ width is inversely proportional to the slope of the dispersion ($\propto$$1/\sqrt{E}$)~\cite{Shirane}.
This leads to excitation at a finite energy $E$ at a $q$ position deviating from the $\Gamma$ point.
The excitation forms a spherical shell in the 3D reciprocal ${\bf Q}$-space.
For the 3D spherical shell, the surface area $\sim 4\pi q^2$ is proportional to the energy for a ferromagnet.
The multiplication of $E$ by $1/\sqrt{E}$ results in $\sqrt{E}$ for the $q$-integrated intensity of a constant-$E$ scan.
The same $E$-dependence of $\chi^{\prime\prime}(E)$ is expected in this YIG around the $\Gamma$ point because of the quadratic dispersion as follows:
\begin{equation}
	\chi^{\prime\prime}(E)=\chi^{\prime\prime}_0 \sqrt{E-E_g},
	\label{eq:6}
\end{equation}
where $\chi^{\prime\prime}_0$ is a constant value.
Although the data in Fig.~\ref{fig:8} were obtained under five different conditions with three spectrometers, all the values follow the same trend below 14~meV, which can be reproduced by Eq.~\ref{eq:6} at $E_g = 0$.
The fitted value of $\chi^{\prime\prime}_0$ below 14~meV in Fig.~\ref{fig:8} was $88\pm 4$  $\rm \mu_B^2$eV$^{-1.5}$Fe$^{-1}$.
On the basis of the good fitting of the MDOS, the validity of the theoretical simple model of LSSE is proved~\cite{Kikkawa2015}.
Under this condition, the MDOS, ${\mathcal D}_M$ in our simple model, can be described by the stiffness constant $D$ at $n(E)=0$.
Moreover, the MDOS is also proportional to the normalized $\chi^{\prime\prime}(E)$ obtained for the lowest-$E$ branch as follows:  
\begin{eqnarray}
{\mathcal D}_M(E)&=&\frac{n_{\rm mode}D^{-3/2}}{(2\pi)^2 40}\sqrt{E-E_g}  \nonumber\\
&=&\frac{A\chi^{\prime\prime}_0}{g^2 \mu_{\rm B}^2 S(S+1)}\sqrt{E-E_g}\;,
\label{eq:7}
\end{eqnarray}
where $A$ is a constant value and 40 is the number of Fe sites in the crystallographic unit cell with a cubic lattice parameter $a=12.36$ \AA.
The value $g^2 \mu_{\rm B}^2 S(S+1)$ is 35 ${\rm \mu_B}^2$Fe$^{-1}$ for Fe$^{3+}$.
However, there are only 20 Fe sites in the magnetic unit cell, where the MDOS is only proportional to the volume per Fe site.
In the calculation by the {\sc spinW} software~\cite{spinw}, 20 Fe sites result in 20 modes in the first Brillouin zone. 
Here, we focus on the lowest-$E$ acoustic magnon mode with $+$ polarization.  

The constant value $A$ is basically unity because of the sum rule for $\chi^{\prime\prime}(E)$ in the $E$-integration at $n(E)=0$. Based on our experimentally obtained stiffness constant of 633~meV\AA$^2$, the constant value $A$ became $0.94\pm 0.02$ at $n_{mode}=1$ (single-mode case).
Thus, we confirmed the single mode for the lowest-$E$ magnon branch from the absolute intensity estimation.

Equation~\ref{eq:7} can be regarded as a Debye model of magnons.
The difference of the constant value from unity may come from our experimental errors and our simplified model. 

Above 14~meV, however, the magnon dispersion deviates from the quadratic function, resulting in the upturn of $\chi^{\prime\prime}(E)$ in Fig.~\ref{fig:8}, which is schematically shown as a dashed line in Fig.~\ref{fig:8}.

What is the meaning of the single mode?
Our result suggests that the mode is only a single polarization, as expected theoretically, although the present inelastic {\it unpolarized} neutron scattering cannot distinguish two polarizations.
This contrasts with doubly degenerate modes in the lowest-$E$ magnon dispersion of an antiferromagnet, which often split in $Q$ due to the Dzyaloshinskii--Moriya interaction~\cite{Park}. 
The two types of polarization modes in YIG are split in energy. In Sect. 4, results from inelastic {\it polarized} neutron scattering will be shown to elucidate the polarization of each magnon mode~\cite{Nambu2020}. 

According to theoretical calculations on YIG~\cite{Barker}, there are 20 modes in the first magnetic Brillouin zone, where 12 modes have $+$ polarization and the other 8 modes have $-$ polarization.
These correspond to $\chi^{\prime\prime}_{yx}$ and $\chi^{\prime\prime}_{xy}$, respectively. The two types of modes in YIG are split in energy corresponding to the energy splitting of up and down spins.
The total number of magnon modes corresponds to the number of Fe in the magnetic unit cell.
This means that the magnetic moment has only one degree of freedom.
On the other hand, a phonon has three degrees of freedom per atom.
They are two transverse modes and one longitudinal mode simultaneously dispersing from the $\Gamma$ point.

Here, we obtained the relation between the magnon dispersion and the dynamical spin susceptibility in the quadratic dispersion case around the $\Gamma$ point.
The validity of the relation may not be limited to only around the $\Gamma$ point nor to the quadratic dispersion.
As far as the dynamic structure factor can be approximated, the dynamical spin susceptibility should be proportional to the MDOS.
Alternatively, if one knows the dynamic structure factor in the $Q$--$E$ space that one wants to study, the present relation can be universally applied.
As an example, the present relation is used to study the ultralow-energy dispersion in the following subsection. 

\subsection{Ultralow-energy spin-wave excitations}

\begin{figure}[t]
	\begin{center}
		\includegraphics[width=0.8\linewidth,clip]{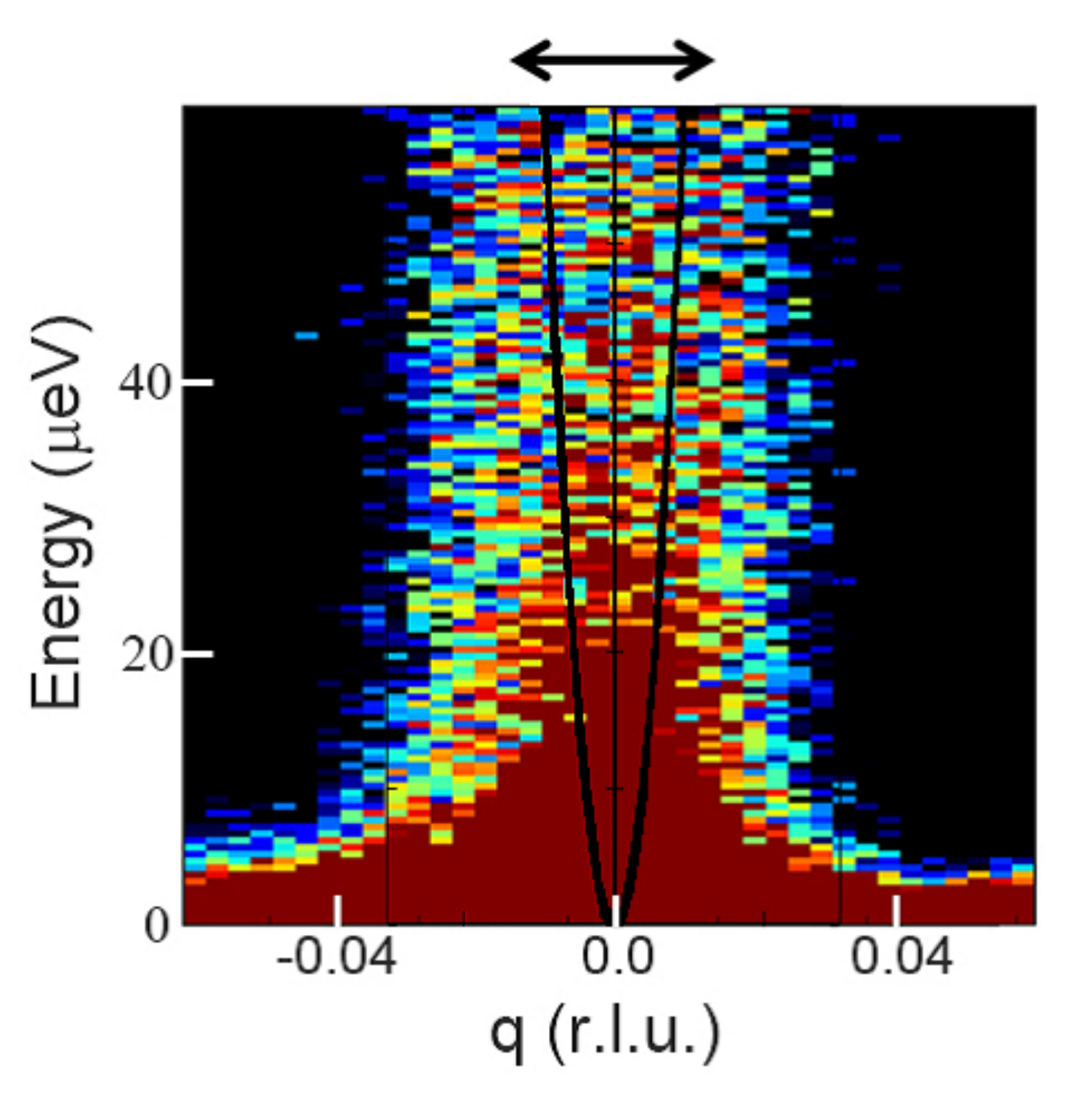}
	\end{center}
	\caption{(Color online) $q$-$E$ spectrum of YIG at $Q = (0, 2, -2)$ measured by BL02 DNA with the expected spin-wave dispersion (solid quadratic line), where $q$ is along [2 -1 -1]$_{\rm cubic}$ (normal to the scattering plane). The instrumental $Q$-resolution, 0.02 \AA$^{-1}$, is shown by the left-right arrow above the figure.} 
	\label{fig:9}
\end{figure}

LSSE depends on the detailed structure of the MDOS, ${\mathcal D}_M$, including the Zeeman energy gap~\cite{Kikkawa2015,Kikkawa2016}.
The Zeeman energy gap in the ferromagnetic dispersion results in the reduction of thermally excited spin current.
Therefore, it is important to determine the Zeeman energy gap directly by INS. 
Ultralow-energy magnon excitation of YIG below 45~${\mu}$eV has been measured, as shown in Fig.~\ref{fig:9} using the inverted-geometry spectrometer BL02 DNA~\cite{Shibata} at J-PARC.
A pulse-shaping chopper with a 3 cm slit rotating at a speed of 225~Hz results in a fine energy-resolution of $3.44\pm 0.02$~$\mu$eV at $E=0$ meV.
Along the vertical $q$-direction, the $Q$-resolution is about 0.02~\AA$^{-1}$.
It is usually difficult to obtain the dispersion at such a low energy by a constant-$E$ or constant-$Q$ scan.
Instead, we estimated the magnon dispersion from the energy dependence of the dynamical spin susceptibility $\chi^{\prime\prime}(E)$ on an absolute scale by assuming the quadratic dispersion of Eq.~\ref{eq:2} described in the previous subsection. 
Here, a magnetic field of about 0.1~T was applied along [111]$_{\rm cubic}$.
This direction is parallel to the growth direction of the rod-shaped crystal, leading to the removal of magnetic domain walls under a magnetic field.

\begin{figure}[t]
	\begin{center}
		\includegraphics[width=0.8\linewidth,clip]{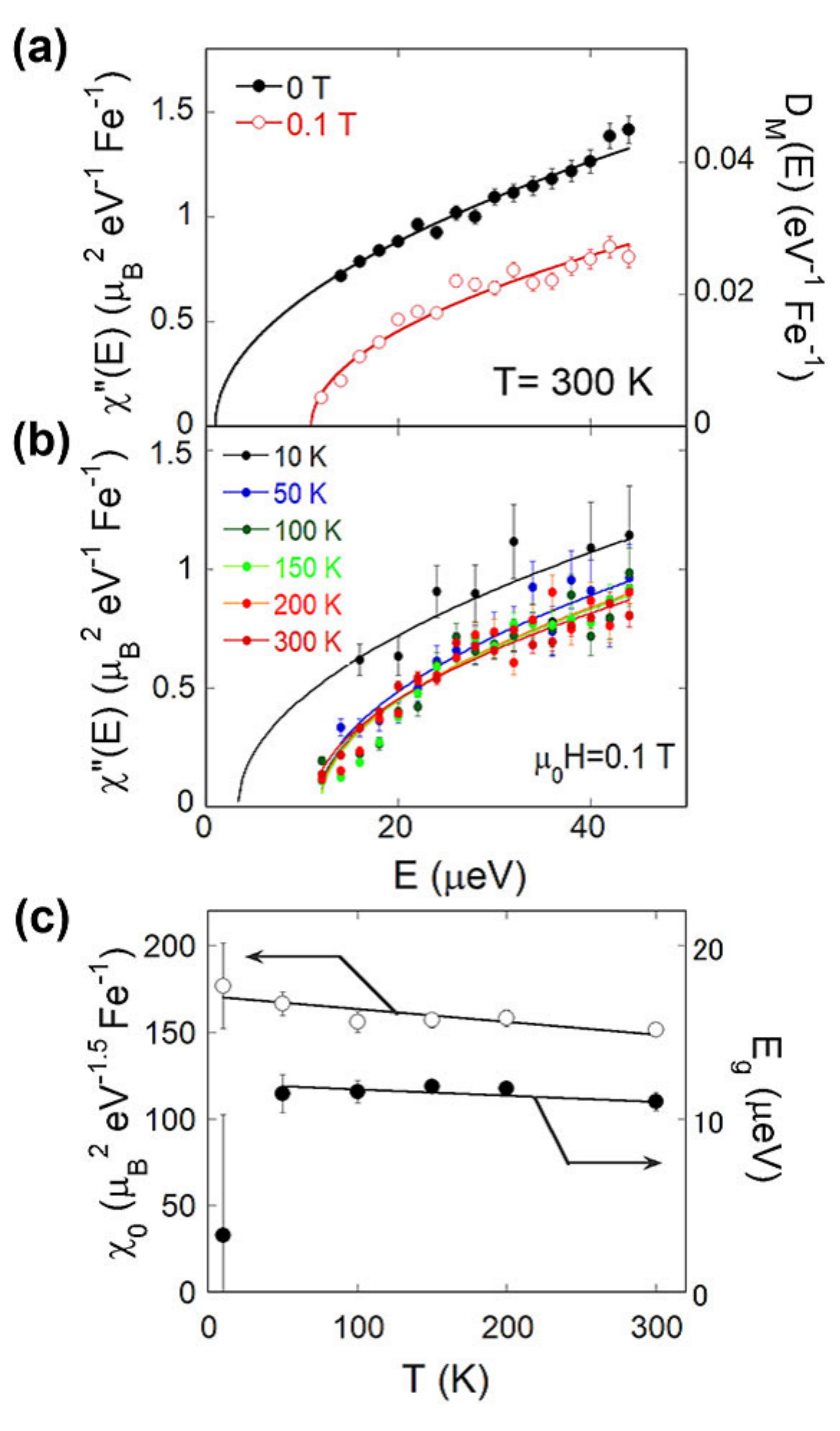}
	\end{center}
	\caption{(Color online) Energy dependence of $Q$-integrated dynamical spin susceptibilities $\chi^{\prime\prime}(E)$. (a) $\chi^{\prime\prime}(E)$ and ${\mathcal D}_M$ at 300~K with (open circles) and without (closed circles) magnetic field. Solid lines are fittings by Eq.~\ref{eq:6}. (b) $\chi^{\prime\prime}(E)$ in the temperature range from 10 to 300~K with a magnetic field of 0.1~T along [111]$_{\rm cubic}$. (c) Temperature dependence of the obtained parameters $\chi^{\prime\prime}_0$ (open circles) and $E_g$ (closed circles) in Eq.~\ref{eq:6}. The solid lines are guides to the eye. Reprinted with permission from Shamoto {\it et al}.~\cite{Shamoto2020} ({\copyright}$\,$ 2020 The American Physical Society).} 
	\label{fig:10}
\end{figure}

In Eq.~\ref{eq:2}, the latter anisotropy term was $0.9\pm 0.5$ $\mu$eV, estimated from the zero-field INS measurement in Fig.~\ref{fig:10}.
Because the total gap was $11.0\pm 0.5$~$\mu$eV under a magnetic field, the Zeeman energy gap becomes $10.1\pm 0.7$~$\mu$eV, suggesting the applied magnetic field $B$ of $0.088\pm 0.006$~T. 
This value agrees with $0.1\pm 0.01$~T measured by a Gauss meter.
The magnetic intensity will weaken under a magnetic field normal to the ${\bf Q}$-vector by a factor of 4/3 due to the angle-dependent term.
The solid lines with the fixed ratio in Fig.~\ref{fig:10}(a) reproduce the intensities very well.
This suggests that the magnetic domain walls are fully removed from the crystal under the magnetic field of 0.1~T, whereas the magnetic domains are randomly oriented under zero magnetic field.
Figure~\ref{fig:10}(b) shows the temperature dependence of the dynamical spin susceptibility $\chi^{\prime\prime}(E)$. The fitted parameters are shown in Fig.~\ref{fig:10}(c).
The Zeeman energy gap nearly closed at 10~K.
The observed intensity at 10~K was weak because of the small Bose factor, resulting in the data scattering of $\chi^{\prime\prime}(E)$ in Fig.~\ref{fig:10}(b).
Because of the large errors at 10~K in Fig.~\ref{fig:10}(c), ambiguity remains in the fitting.
This gap closure at 10~K is confirmed by specific heat measurement under a magnetic field along [111]$_{\rm cubic}$ as discussed in the next subsection.
From Eq.~\ref{eq:7} with $n_{mode}=1$ and $A=1$, the magnon stiffness constant $D$ becomes
\begin{equation}
D=\left\{\frac{g^2 \mu_{\rm B}^2 S(S+1)}{{(2\pi)^2 40}\chi^{\prime\prime}_0}\right\}^{2/3},
\label{eq:8}
\end{equation}
where 40 is the number of Fe sites in the crystal unit cell.

In Fig.~\ref{fig:10}(c), $\chi^{\prime\prime}_0$ increases from $157.0\pm 4.3$~${\rm \mu_B}^2$eV$^{-3/2}$Fe$^{-1}$ at $T=150$~K to $177\pm 25$~${\rm \mu_B}^2$eV$^{-3/2}$Fe$^{-1}$ at 10~K. The enhancement corresponds to about 13\%.
This has two possible explanations.
One is the magnon softening with decreasing temperature.
The other is the spin canting observed in the magnetization, which increases the INS intensity via the angle-dependent term. 
The stiffness constant $Da^2$ estimated from $\chi^{\prime\prime}_0$ decreases from $415\pm 7$~meV\AA$^2$ at 150~K to $383\pm 76$~meV\AA$^2$ at 10~K.
The value at 10~K is smaller than our previous result of $633\pm 17$~meV\AA$^2$ at $\sim$20~K obtained from the magnon dispersion in the energy range from 2 to 14~meV~\cite{Shamoto2018}.
The present result is opposite to the standard expectation that a magnon is hardened with decreasing temperature.
The softening is also observed below 150~K via the microwave spin-wave resonance of YIG under a magnetic field along [111]$_{\rm cubic}$~\cite{LeCraw}.
The stiffness constant is estimated assuming that a relevant phonon velocity is constant.
The stiffness constant $D$ decreases by about 5\% due to the magnon softening, which is well reproduced by the random phase approximation with sublattice magnetizations $M_a$ and $M_d$ at the $a$ and $d$ sites, respectively.
$D$ is not proportional to the total magnetization $M$ but is expressed in terms of $M_a$ and $M_d$ as follows~\cite{LeCraw}:
\begin{equation}
D=B\frac{-8J_{aa}M_{a}^{2}-3J_{dd}M_{d}^{2}+5J_{ad}M_{a}M_{d}}{3M_{d}-2M_{a}},
\label{eq:9}
\end{equation}
where $B$ is a constant value.
This magnon softening corresponds to an 8\% increase in $\chi^{\prime\prime}_0$.
This is not large enough for the enhancement of about 13\%.     
Although our magnetic structure refinement at about 295 K does not show any spin canting, the result suggests that the spins cant at low temperatures below 150~K, as discussed in the next subsection.

\subsection{Magnetization and specific heat results}

\begin{figure}[t]
	\begin{center}
		\includegraphics[width=0.6\linewidth,clip]{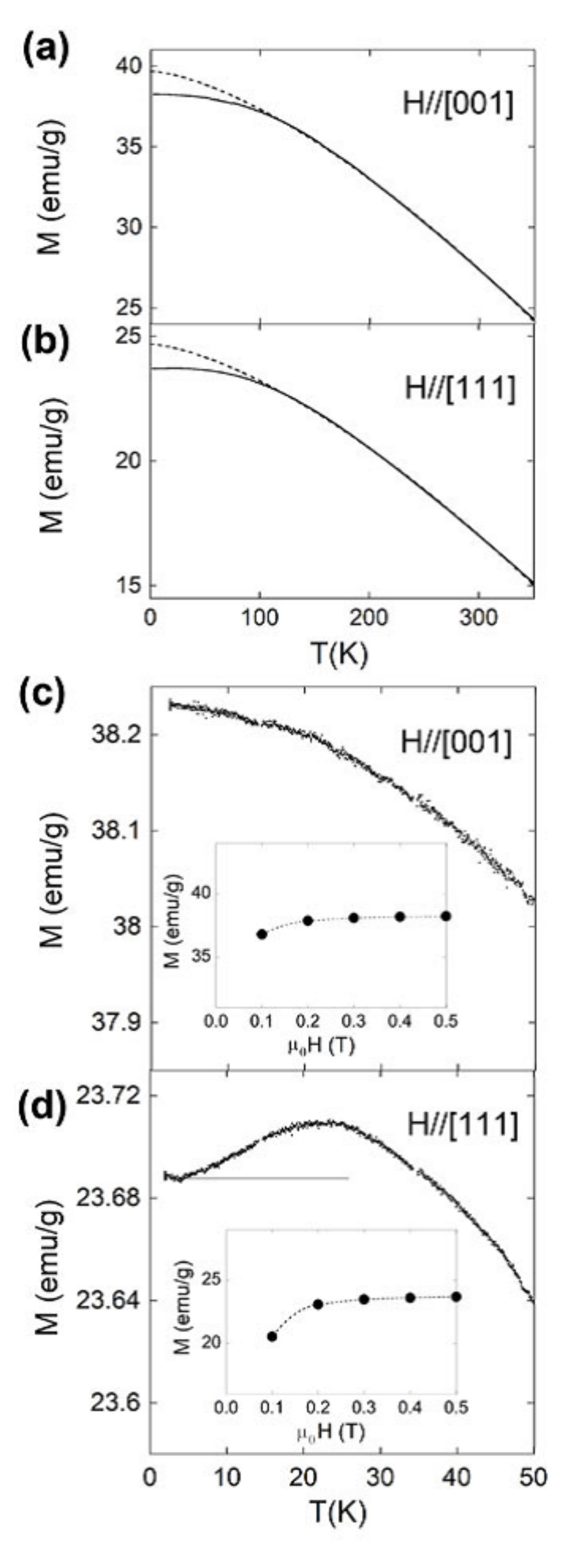}
	\end{center}
	\caption{Magnetization as a function of temperature. Observed magnetization under a magnetic field of 0.5~T along [001]$_{\rm cubic}$ (a) and [111]$_{\rm cubic}$ (b) (solid lines) with Bloch-type magnetization (broken lines) fitted in the temperature range from 150 to 350~K. Enlarged temperature dependences of the magnetization at $B=0.5$~T along [001]$_{\rm cubic}$ (c) and [111]$_{\rm cubic}$ (d). The insets show the magnetic field dependence of the magnetization. The thin horizontal line in (d) is a guide to the eye. Reprinted with permission from Shamoto {\it et al}.~\cite{Shamoto2020} ({\copyright}$\,$ 2020 The American Physical Society).}
	\label{fig:11}
\end{figure}

The observed magnetic anomaly below 150~K was examined by investigating the magnetization and specific heat of YIG.
The results will be compared with INS results.
Figures~\ref{fig:11}(a) and \ref{fig:11}(b) show the temperature dependence of magnetization under a magnetic field cooling of 0.5~T along [001]$_{\rm cubic}$ and [111]$_{\rm cubic}$, respectively.
For these measurements, non-cylindrical small crystals were used due to the limitation of these measurements.
A magnetic field of 0.5~T, larger than 0.1~T, was applied in Fig.~\ref{fig:11}, because of the crystal shapes having the ambiguity of the demagnetization effect.
To confirm the difference by the demagnetization, the magnetic field dependence was measured from 0.1 to 0.5~T as shown in the insets of Figs.~\ref{fig:11}(c) and \ref{fig:11}(d).
They show that 0.1~T is not large enough for the full magnetization of YIG owing to the small demagnetization effect of the crystals.
The magnetization at low temperatures is usually decreased by the magnon excitation with increasing temperature, where the temperature dependence is expressed by the Bloch $T^{3/2}$ rule~\cite{Kikkawa2015}.
In the present case, however, the Bloch rule can be applied to the magnetization only above 150~K as shown in Figs.~\ref{fig:11}(a) and \ref{fig:11}(b).
The fittings lead to similar values of $\zeta=5.96$--$7$ $\times$ 10$^{-5}$~K$^{-3/2}$ for $M=M_0(1-\zeta T^{3/2})$.
They are slightly larger than the reported values of 5.20--5.83 $\times$ 10$^{-5}$~K$^{-3/2}$ in the previous study~\cite{Kikkawa2015}.  
Below 150~K, the magnetization under a magnetic field along [001]$_{\rm cubic}$ shows a continuous increase with decreasing temperature (Fig.~\ref{fig:11}(c)). On the other hand, a peak appears at approximately 25~K in the magnetic field along [111]$_{\rm cubic}$ (Fig.~\ref{fig:11}(d)), suggesting a crossover.

\begin{figure}[t]
	\begin{center}
		\includegraphics[width=0.8\linewidth,clip]{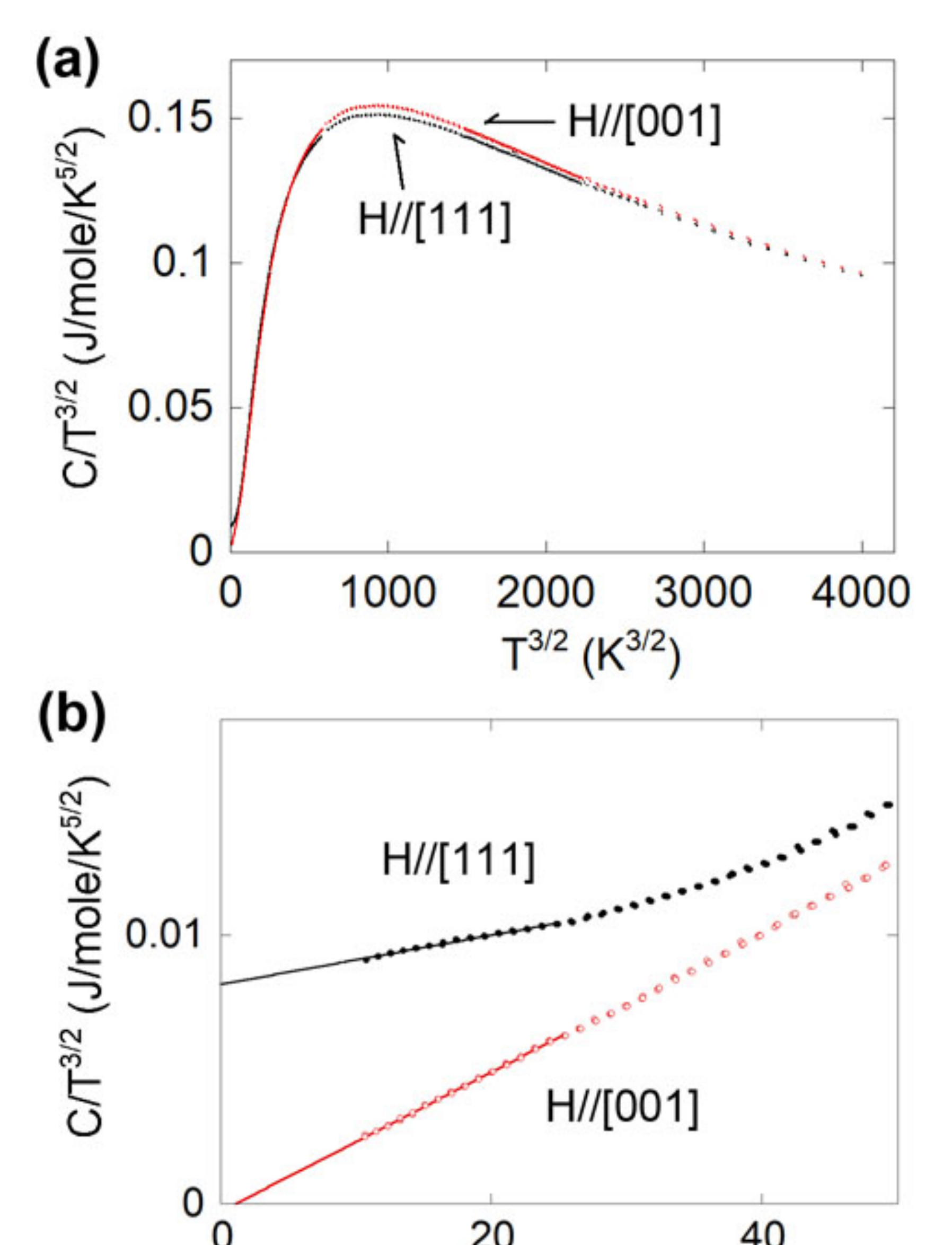}
	\end{center}
	\caption{(Color online) Specific heat capacity $C/T^{3/2}$ plotted as a function of $T^{3/2}$. (a) The magnetic field directions are parallel to [001]$_{\rm cubic}$ (red dots) and [111]$_{\rm cubic}$ (black dots) under 1 T. Both heat capacities are plotted in a wide temperature range up to 250 K. The upper values (red dots) around the peak at $T$= 96~K ($\sim$920 K$^{3/2}$) are measured at $B$=1 T along [001]$_{\rm cubic}$, which at low temperatures become lower than those (black dots) measured at $B$=1 T along [111]$_{\rm cubic}$. (b) Enlarged temperature dependence of the heat capacities for [001]$_{\rm cubic}$ (red open circles) and [111]$_{\rm cubic}$ (black closed circles) magnetic fields at low temperatures below 14~K. The difference becomes large below $T$= 9 K ($\sim$25 K$^{3/2}$). The linear solid lines are fitted results within the range, extrapolated to the ends. Reprinted with permission from Shamoto {\it et al}.\cite{Shamoto2020} ({\copyright}$\,$ 2020 The American Physical Society).
	}
	\label{fig:12} 
\end{figure}

Specific heat measurement has also been performed to observe the anomalous behavior at 25~K as shown in Figs.~\ref{fig:12}(a) and \ref{fig:12}(b).
A 3D ferromagnet/ferrimagnet magnon without a gap is expected to show a specific heat proportional to $T^{3/2}$, in addition to the phonon term of $T^3$ at low temperatures~\cite{Kittel} as follows:
\begin{equation}
\frac{C}{T^{3/2}}=A+BT^{3/2},
\label{eq:8}
\end{equation}
where $A=0.113k_{\rm B}(Da^2 /k_{\rm B})^{-3/2}$ and $B=12\pi^{4}N_{\rm A} k_{\rm B}/(5\theta_{\rm D}^{3})$ with $k_{\rm B}$ the Boltzmann constant, $N_{\rm A}$ the Avogadro number for the formula weight (Y$_3$Fe$_5$O$_{12}$), and $\theta_{\rm D}$ the Debye temperature. 

In Figs.~\ref{fig:12}(a) and \ref{fig:12}(b), $C/T^{3/2}$ is plotted as a function of $T^{3/2}$.
Figure~\ref{fig:12}(a) shows the specific heat over a wide temperature range.
The specific heat capacities under the two magnetic field directions start to deviate from each other below around 150~K, suggesting that the magnon anomaly may have already started by 150~K.
This corresponds to the non-Bloch-type temperature dependence of magnetization below 150~K.
The anomaly corresponds to the magnon softening observed below 150 K by microwave spin-wave resonance~\cite{LeCraw}.
In addition, the first-order anisotropy constant anomalously increases below 150~K in a ferromagnetic resonance study of YIG~\cite{Dillon}, suggesting different origins from a simple sublattice magnetization effect such as spin canting.
Figure~\ref{fig:12}(b) shows the same data at temperatures below 14~K.
The slope corresponds to the Debye temperature $\theta_{\rm D}$, whereas the extrapolation to the $y$-axis leads to the stiffness constant $Da^2$, which is expected to be common for all the plots in Fig.~\ref{fig:12}(b).
However, both parameters below 9~K ($\sim$25~$K^{3/2}$) are different between the two conditions.
In the case of $B$ along [001]$_{\rm cubic}$, $A$ becomes $-2.8\pm 0.2\times 10^{-4}$~J/mole/K$^{5/2}$, whereas $A$ is $8.17\pm 0.02\times 10^{-3}$~J/mole/K$^{5/2}$ at $B$ along [111]$_{\rm cubic}$. 
The former negative $A$ value can be regarded as an artifact coming from the assumed magnon excitation without a gap in the narrow temperature range from 5.0 to 8.8~K.
The latter finite magnon contribution to the specific heat suggests that the magnon energy gap disappears under $B$ along [111]$_{\rm cubic}$.

The stiffness constant $Da^2$ obtained from the $A$ value at the magnetic field along [111]$_{\rm cubic}$ was 312~meV\AA$^2$.
This value is fairly consistent with $383\pm 76$~meV\AA$^2$ obtained from the ultralow-energy magnon at 10~K.
However, these values are much smaller than our previous result of $633\pm 17$~meV\AA$^2$ estimated at $\sim$20~K.

In the magnetic field along [001]$_{\rm cubic}$, the Debye temperature was 195.8~K, whereas it became 277.0~K in the magnetic field along [111]$_{\rm cubic}$. The Debye temperature seems to change depending on the magnetic field direction. This lattice hardening can be seen below 9~K in Fig.~\ref{fig:12}(b).

We did not observe any peaks in the temperature dependence of specific heat in the whole range of the measured temperature, suggesting no appreciable phase transition.
This suggests that this anomaly may originate from a crossover from a ferrimagnet to a canted ferrimagnet. 

We observed a new crossover at 25~K below the precursor anomaly at 150~K.
Regarding the precursor anomaly, the deviation from the Bloch rule below 150~K in Fig.~\ref{fig:11}(a) reaches 3.6\% (${\bf H}\parallel$[001]$_{\rm cubic}$) and 4.0\% (${\bf H}\parallel$[111]$_{\rm cubic}$) at the lowest temperature of 2.5~K.
The amount of suppression corresponds to about 15.4 and 16.3 degrees in the canting angles of magnetic moments, respectively.
The spin canting is expected to increase the INS intensity due to the angle-dependent term $\{1+(\hat{\tau} \cdot \hat{\eta})^2\}_{av}$ in Eq.~\ref{eq:5}.
The canting of 15.4 degrees leads to a 7\% enhancement of the intensity, which is smaller than the enhancement of about 13\% in Fig. \ref{fig:10}(c).
However, the remaining part can be explained by the magnon softening based on the microwave spin resonance result.~\cite{LeCraw}
The summation of these two components is about 15\%, which is fairly close to the total enhancement of about 13\%.

Regardning the magnetization peak at 25~K in Fig.~\ref{fig:11}(d), this magnetization anomaly accompanies the Zeeman energy-gap closure.
The dielectric relaxation anomaly also exhibits the same temperature and the same magnetic field direction dependence~\cite{Yamasaki2009}, suggesting the same origin as the magnetization anomaly.
In this temperature range, the lattice must strongly couple with the spin via spin-orbit coupling.

The crystal structure of YIG is trigonally distorted at room temperature~\cite{Shamoto2018}.
The symmetry axis [111]$_{\rm cubic}$ coincides with the magnetic field direction along [111]$_{\rm cubic}$, which is also the magnetic easy axis. However, it is difficult to understand why the dielectric property appears at low temperatures as a crossover transition. 
One possible scenario is that thermally activated itinerant Fe$^{2+}$ impurities are frozen out at low temperatures~\cite{Yamasaki2009}.
Impurity Fe$^{2+}$ centers provide large spin-orbit coupling effects~\cite{Kohara}.
Therefore, the observed magnetic anomaly may appear through the spin-orbit coupling effect due to the localization of Fe$^{2+}$ centers below 150~K.

Based on the magnon anomaly appearing under a magnetic field along [111]$_{\rm cubic}$, we may expect the low-temperature spin Seebeck effect to reflect this anomaly.
So far, no experiments under this condition have been reported.

It is necessary to investigate how the spin Seebeck effect changes depending on the magnetic field direction relative to the crystal axis.
Although the $\chi^{\prime\prime}_{xy}$ polarization mode may change the magnon character with the spin canting in the low-lying acoustic magnon mode, the Zeeman energy gap closure can increase the spin current, especially at low temperatures.
This type of experiment may provide further information about the enhancement of the spin Seebeck effect.

The relationship between the magnon dispersion and the dynamical spin susceptibility $\chi^{\prime\prime}(E)$ has been useful for investigating ultralow-energy magnons beyond the $Q$-resolution.
This method was indispensable for the study of ultralow-energy magnons.
By this method, the Zeeman energy gap anomaly was found under a magnetic field along [111]$_{\rm cubic}$ in YIG.
Magnetization measurement also revealed a similar anomaly under a magnetic field along the same direction.
Moreover, specific heat measurement also revealed that the Zeeman energy gap closes at low temperatures.
The increase of $\chi^{\prime\prime}_0$ was discussed in terms of the following two possibilities.
One is that the sublattice magnetization effect softens the ultralow-energy magnon dispersion.
The other is that spin canting emerges at low temperatures.
The former magnon softening estimated from the previous report\cite{LeCraw} was an increase of about 8\% in $\chi^{\prime\prime}_0$, whereas the latter spin canting of 15.4 degrees estimated from the magnetization contributes to an increase of about 7\%. 
The large increase of $\chi^{\prime\prime}_0$ in Fig. \ref{fig:10}(c) was roughly explained by these two effects.

The spin canting anomaly was attributed to the magnetic crossover transition in YIG below 150~K.
Although this still remains as a speculation because of the large error bars in $\chi^{\prime\prime}_0$ at $T$= 10~K, this spin canting anomaly may play an important role in enhancing spintronic properties, such as the spin Seebeck and ultrasound spin-pumping effects.

\section{Inelastic $\bm{Polarized}$ Neutron Scattering}

\subsection{Spin current carrier in insulators}

\begin{figure}[t]
	\begin{center}
		\includegraphics[width=0.8\linewidth,clip]{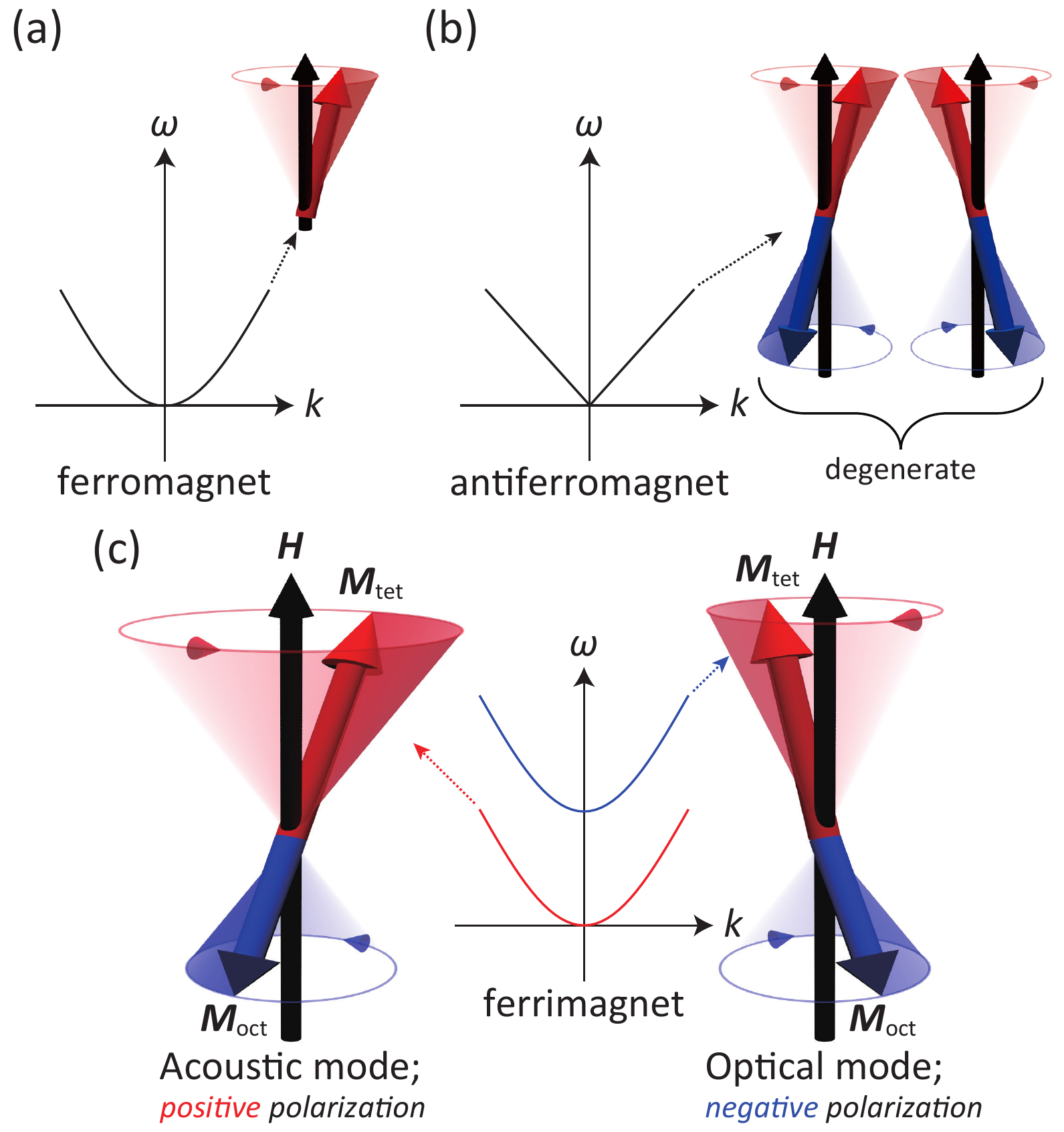}
	\end{center}
	\caption{(Color online) Illustration of magnon polarization for (a) a ferromagnet, (b) an antiferromagnet, and (c) two magnon modes in a ferrimagnet. The ``positive'' polarization acoustic mode is a coherent right-handed circular precession of the moments, whereas the ``negative'' polarization optical mode is a left-handed precession dominated by the exchange interaction between Fe$_{\rm oct}$ and Fe$_{\rm tet}$ sites. Reprinted with permission from Nambu {\it et al}.~\cite{Nambu2020} ({\copyright} 2020 The American Physical Society).}
	\label{nambu-fig1} 
\end{figure}

The spin current in insulators can be carried by the precessional motion of the ordered moments.
More precisely, the thermal spin motive force, i.e., ``spin pumping,'' which governs the spin Seebeck signal, is proportional to the product of the integrated energy and the chiral correlation function (transverse component of the magnons)~\cite{Xiao2010}.
We commence this section with a review of the direction of the precessional motion, i.e., magnon polarization~\cite{Nambu2020}.
The magnon polarization had never been directly measured in any material.
According to the Landau--Lifshitz equation without the Gilbert damping term for simplicity,
\begin{align}
	\frac{\partial\vec{M}}{\partial t}=-\gamma\vec{M}\times\vec{H}^{\rm eff},
	\label{LL}
\end{align}
a magnetic moment precesses counterclockwise around the effective magnetic field direction, where $\gamma$ is the gyromagnetic ratio of the moments.
This motion can be defined as ``positively'' polarized.
The collective excitations in single-domain ferromagnets also precess only counterclockwise; hence, all ferromagnetic magnons have a positive polarization (Fig.~\ref{nambu-fig1}(a)).
Simple collinear antiferromagnets have two magnon modes with opposite polarization (Fig.~\ref{nambu-fig1}(b)), but these are degenerate unless large magnetic fields or Dzyaloshinskii--Moriya interactions are turned on.
Simple ferrimagnets have two anti-aligned sublattices and also support two magnon polarizations, but the inter-sublattice exchange field naturally separates the branches of opposite polarization into acoustic and optical modes (Fig.~\ref{nambu-fig1}(c)).
The energy gap between these modes can be large; hence, spectroscopic studies have the potential to observe this character.
YIG is a good material for attempting such measurements, since it is well studied and frequently employed for spintronics and magnonics.

The magnon polarization in ferrimagnets for the uniform ($Q=0$) modes can be understood with Eq.~\ref{LL}.
With $j\in [{\rm tet}, {\rm oct}]$, the effective magnetic field can be written as~\cite{Schlomann1960}
\begin{align}
	\vec{H}_j^{\rm eff}&=-\frac{\partial U}{\partial \vec{M}_j}\nonumber\\
	&=\frac{\partial}{\partial \vec{M}_j}\left(\frac{1}{2}\Lambda_{\rm tet}M_{\rm tet}^2+\frac{1}{2}\Lambda_{\rm oct}M_{\rm oct}^2-\Lambda\vec{M}_{\rm tet}\cdot\vec{M}_{\rm oct}\right),
\end{align}
with $\Lambda$ being exchange interaction constants.
We respectively divide $\vec{M}_j$ and $\vec{H}_j^{\rm eff}$ into static and dynamic parts, $\vec{M}_j(t)=\vec{M}_{j,0}+\vec{m}_j(t)$ and $\vec{H}_j^{\rm eff}(t)=\vec{H}_{j,0}^{\rm eff}+\vec{h}_j^{\rm eff}(t)$.
After linearization, choosing the quantization axis along $z$, $M_{\rm tet}^z=M_{\rm tet}$, $M_{\rm oct}^z=-M_{\rm oct}$, and $m^{\pm}_j=m^x_j\pm im^y_j$, the secular equation reads
\begin{equation}
	\begin{pmatrix}
		\pm\omega-\gamma_{\rm tet}\Lambda M_{\rm oct} & -\gamma_{\rm tet}\Lambda M_{\rm tet}\\
		\gamma_{\rm oct}\Lambda M_{\rm oct} & \pm\omega+\gamma_{\rm oct}\Lambda M_{\rm tet}
	\end{pmatrix}
	\begin{pmatrix}
		m^{\pm}_{\rm tet}\\
		m^{\pm}_{\rm oct}
	\end{pmatrix}
	=0.
\end{equation}
One of the eigenenergies is $\omega_{\rm acoustic}=0$ for the acoustic mode, whereas the other gives $\omega_{\rm optical}=\Lambda\left(\gamma _{\rm oct}M_{\rm tet}-\gamma_{\rm tet}M_{\rm oct}\right)$ corresponding to the optical gap.
Eigenoscillations within the $xy$-plane can then be easily calculated, where the precession radii give $m^{\pm}_{\rm tet}/m^{\pm}_{\rm oct}=-M_{\rm tet}/M_{\rm oct}$ and $-\gamma_{\rm tet}/\gamma_{\rm oct}$ for the acoustic and optical modes, respectively.
The corresponding eigenoscillations are schematically depicted in Fig.~\ref{nambu-fig1}(c): the acoustic mode is a coherent right-handed circular precession in which $\vec{M}_{\rm tet}$ and $\vec{M}_{\rm oct}$ are exactly antiparallel, whereas the optical mode is a left-handed circular precession with a finite canting angle.

\subsection{Polarized neutron scattering cross sections}

Polarized neutron scattering--taking into account the neutrons' spin degree of freedom in the scattering process--gives detailed information of the magnetism~\cite{Chatterji2006}.
It has mainly been used to disentangle the magnetic and nuclear contributions to scattering cross sections~\cite{Moon1969}.
The elucidation of the magnetic moment directions, as well as the symmetry of magnetic fluctuations~\cite{Kakurai1984}, has also been demonstrated.
More recently, the ``chiral term''~\cite{Maleyev1995} was used to measure the chiral (handedness of) magnetic order~\cite{Loire2011} and excitations in paramagnetic~\cite{Roessli2002} and magnetically ordered phases~\cite{Lorenzo2007a}.
The chirality observed in these studies is a spatial variation of the {\it non-collinear} magnetic moments caused by effects such as geometrical frustration and the Dzyaloshinskii--Moriya interactions.
Here, we aim for a different property: the intrinsic polarization of the magnetic excitations in a {\it collinear} magnet.
The nuclear-magnetic ``interference term'' has also attracted much attention recently in the context of magnetoelastic coupling.

\begin{figure}[t]
	\begin{center}
		\includegraphics[width=0.8\linewidth,clip]{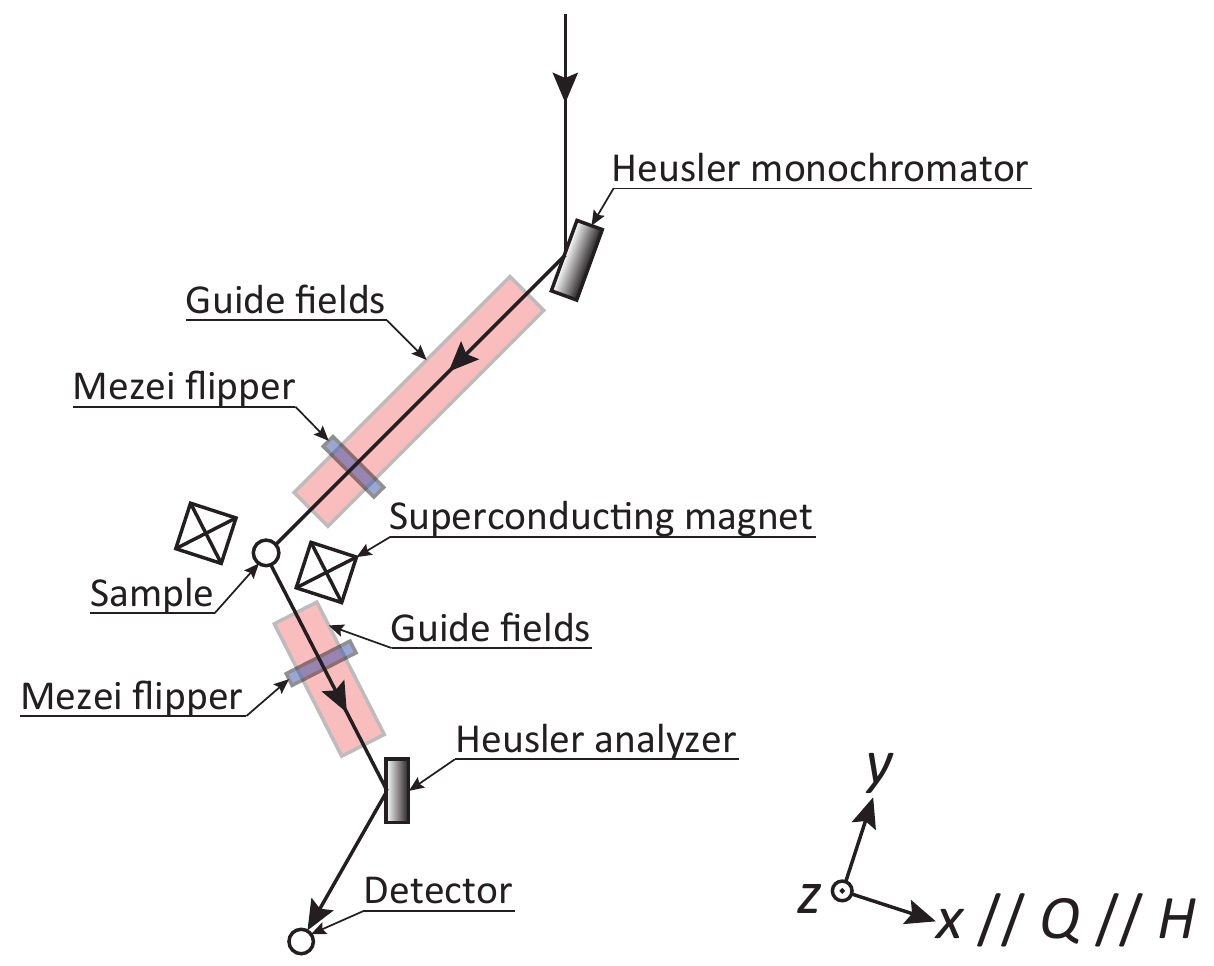}
	\end{center}
	\caption{(Color online) Sketch of the $P_x$-polarized neutron scattering experiment on the IN20~\cite{IN20} instrument at ILL, France, with bold black arrows denoting the neutron path. The scattering coordinate $(x,y,z)$ is also given. Reprinted with permission from Nambu {\it et al}.~\cite{Nambu2020} ({\copyright} 2020 The American Physical Society).}
	\label{nambu-fig2} 
\end{figure}

To detect the magnon polarization, a special coordination of the neutron polarization is required.
The neutron polarization direction can be taken as an arbitrary direction, i.e., it is not restricted only to perpendicular to the scattering plane.
For a schematic understanding, we adopt a simple orthogonal scattering coordinate $(x,y,z)$ ({\it aka} Blume--Maleyev coordination, as in Fig.~\ref{nambu-fig2}), where $x\parallel \vec{Q}$, $y\perp \vec{Q}$, and $z$ is perpendicular to the horizontal scattering plane with the scattering wave vector $\vec{Q}$.
The observations for each neutron polarization direction are summarized via the following cross-section formulae:~\cite{Blume1963,Maleyev1963},
\begin{align}
	&\sigma_x^{\pm\pm}\propto N,						\label{xnsf}\\
	&\sigma_x^{\pm\mp}\propto M_y + M_z \mp M_{\rm ch},	\label{xsf}\\
	&\sigma_y^{\pm\pm}\propto N+M_y\pm R_y,				\label{ynsf}\\
	&\sigma_y^{\pm\mp}\propto M_z,						\label{ysf}\\
	&\sigma_z^{\pm\pm}\propto N+M_z\pm R_z,				\label{znsf}\\
	&\sigma_z^{\pm\mp}\propto M_y,						\label{zsf}
\end{align}
where the ideal case with perfect performances of neutron spin polarizers and flippers is assumed.
The term $\sigma_{\alpha}^{io}$ ($\alpha = x,y,z$) stands for the partial differential scattering cross section (${\rm d}^2\sigma/({\rm d}\Omega{\rm d}E)^{io}$) with $i$ incoming and $o$ outgoing neutrons with $+/-$ neutron polarization.
The nonmagnetic nuclear ($N=\langle N_{Q}N_{Q}^{\dagger }\rangle _{\omega}$), in-plane magnetic ($M_y=\langle M_{Qy}M_{Qy}^{\dagger}\rangle _{\omega}$), out-of-plane magnetic ($M_z=\langle M_{Qz}M_{Qz}^{\dagger}\rangle _{\omega}$), chiral ($M_{\rm ch}=i(\langle M_{Qy}M_{Qz}^{\dagger}\rangle_{\omega}-\langle M_{Qz}M_{Qy}^{\dagger}\rangle_{\omega})$), and interference ($R_{\beta}=\langle N_Q M_{Q\beta}^{\dagger}\rangle_{\omega}+\langle M_{Q\beta}N_Q^{\dagger}\rangle_{\omega}$) ($\beta = y,z$) terms are included.
$\langle N_{Q}N_{Q}^{\dagger }\rangle _{\omega }$ and $\langle M_{Q\beta}M_{Q\beta}^{\dagger}\rangle_{\omega}$ are the spatiotemporal Fourier transforms of the nuclear-nuclear and spin-spin correlation functions, respectively.
The chiral term defines the antisymmetric correlation function within the $yz$-plane, and the interference terms describe the symmetric part of the nuclear-magnetic interference.

Detection of the magnon polarization owing to the chiral term is difficult due to the low scattering intensities.
The chiral terms can only be measured when the applied field and magnetization are aligned with the scattering wave vector $\vec{Q}$.
Magnetic neutron scattering can only detect the spin components perpendicular to this $\vec{Q}$, and these projections in {\it collinear} magnets are considered to be tiny.
Moreover, the signal is contaminated by imperfections in the polarizers and flippers, which are needed to select the incident and scattered neutrons (see Fig.~\ref{nambu-fig2}).
Derived analytical formulae for corrections of the neutron polarization are summarized elsewhere.~\cite{Nambu-arXiv}

\subsection{Mode-resolved magnon polarization}

Single-crystalline samples of YIG were grown by the traveling solvent floating zone method, and the measured Curie temperature $T_{\rm C}=553$~K agrees well with previous reports.
Inelastic polarized neutron scattering data were obtained using the thermal neutron triple-axis spectrometer IN20~\cite{IN20} at the Institut Laue-Langevin, France.
IN20 is equipped with a Heusler (111) monochromator and analyzer to polarize the incident neutron beam and analyze the scattered neutron polarization; the horizontally variable and vertically fixed curvature enables a large polarized neutron flux.
We used a graphite filter in the outgoing beam to suppress higher-order contamination and fixed the final wavenumbers of $k_{\rm f}=2.662$ and 4.1~{\AA}$^{-1}$, corresponding to final energies of $E_{\rm f}=14.7$ and $34.8$~meV, respectively.

A YIG single crystal ($\sim$8~g) was oriented with $(HHL)$ in the horizontal scattering plane, and put into a cryomagnet supplying a horizontal magnetic field of 0.3~T parallel to the momentum transfer $\vec{Q}$ and temperatures between 10 and 300~K.
YIG is a very soft magnet~\cite{Yamasaki2009}, and the external field of 0.3~T was sufficient to fully saturate the magnetization into a single magnetic domain.
The obtained fields were homogeneous over a large area around the sample position and were discontinuously connected to the guide fields along the neutron path.

In IN20, the scattered neutrons are recorded in four {\it channels}: $I_{x}^{++}$, $I_{x}^{--}$, $I_{x}^{+-}$, $I_{x}^{-+},$ where $I_{x}^{io}\ \left(\propto {\rm d}^2\sigma/({\rm d}\Omega{\rm d}E)^{io}\right)$ is the intensity of $i$ incoming and $o$ outgoing neutrons with $+/-$ neutron polarization~\cite{Chatterji2006}.
From the four channels, the non-magnetic nuclear ($N$), magnetic ($M=M_{y}+M_{z}$), and chiral ($M_{\rm ch}$) contributions can be extracted through the following combinations:
\begin{align}
	&N = \langle N_Q N_Q^{\dagger}\rangle_{\omega} = \frac{1}{2}(I_x^{++} + I_x^{--}),\\
	&M = \langle M_{Qy}M_{Qy}^{\dagger}\rangle_{\omega} + \langle M_{Qz}M_{Qz}^{\dagger}\rangle_{\omega} = \frac{1}{2}(I_x^{+-} + I_x^{-+}),\\
	&M_{\rm ch} = i(\langle M_{Qy}M_{Qz}^{\dagger}\rangle_{\omega}-\langle M_{Qz}M_{Qy}^{\dagger}\rangle_{\omega}) = \frac{1}{2}(I_x^{+-} - I_x^{-+}).
\end{align}
Phonon and magnon scatterings in YIG have been separated in terms of the nuclear and magnetic spectra~\cite{Plant1977,Princep2017a,Shamoto2018,Shamoto2020}.
The chiral contribution $M_{\rm ch}$ contains new information about the magnon polarization.

\begin{figure}[t]
	\begin{center}
		\includegraphics[width=0.8\linewidth,clip]{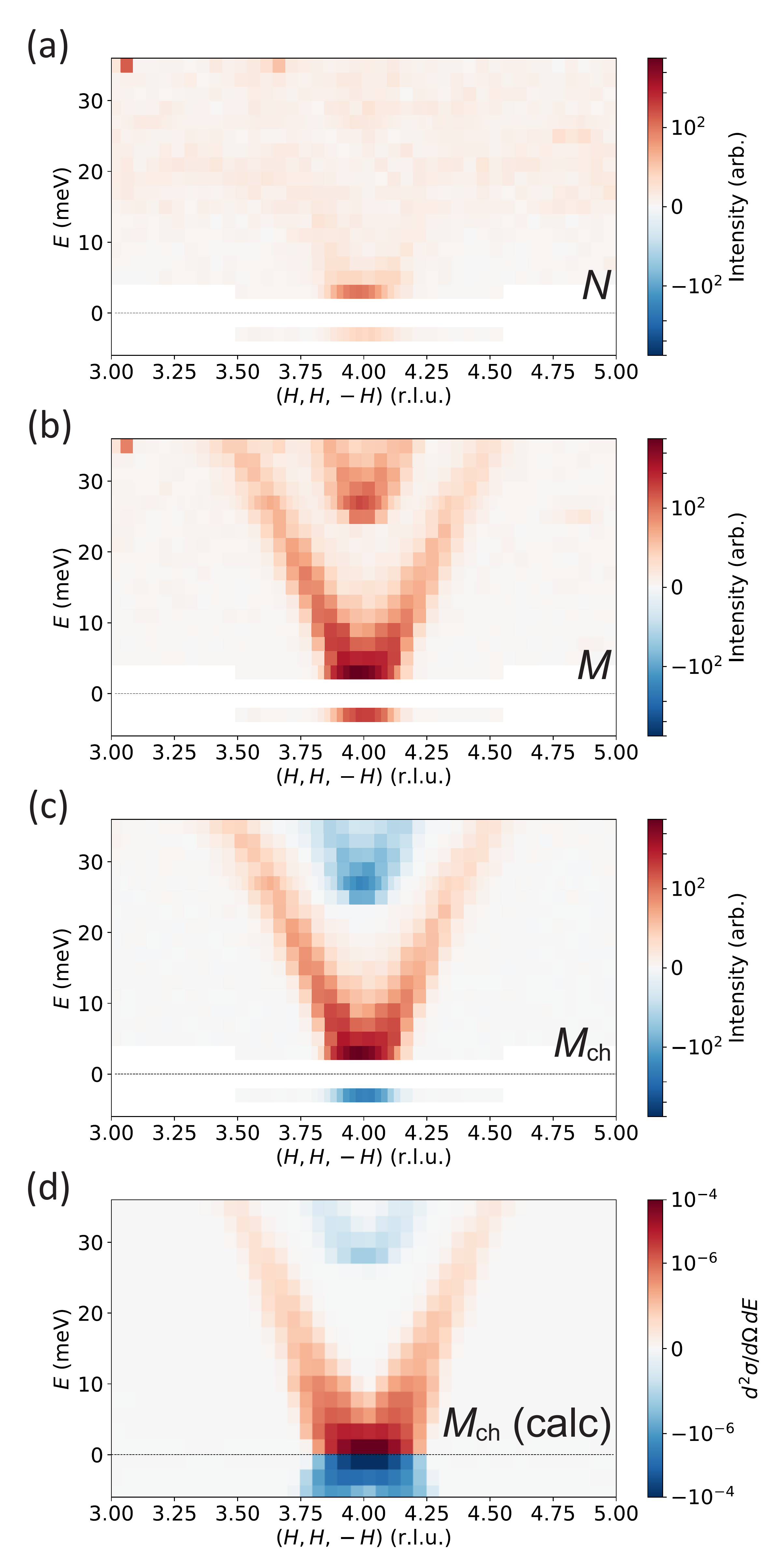}
	\end{center}
	\caption{(Color online) Derived spectra of (a) the nuclear term $N=\frac{1}{2}(I_x^{++}+I_x^{--})$, (b) magnetic term $M=M_y+M_z=\frac{1}{2}(I_x^{+-}+I_x^{-+})$, and (c) chiral term $M_{\rm ch}=\frac{1}{2}(I_x^{+-}-I_x^{-+})$ from mesh scans taken at 293~K. Note that some scans miss the $I^{++}$ channel, which is approximated by $I^{--}$. The chiral term is compared with (d) the calculated resolution-convoluted partial differential scattering cross section. Reprinted with permission from Nambu {\it et al}.~\cite{Nambu2020} ({\copyright} 2020 The American Physical Society).}
	\label{nambu-fig3} 
\end{figure}

Figure~\ref{nambu-fig3} summarizes the results of accumulating many scans and their derivations around $Q=(4,4,-4)$.
The nuclear response is very weak (Fig.~\ref{nambu-fig3}(a)) as intended: the $(4,4,-4)$ intensity is four orders of magnitude smaller than that of the strongest nuclear Bragg peak $(0,0,4)$.
Very weak phonon excitations and/or the imperfections of the neutron polarization and flippers may cause the remaining weak signals. 
The disentangled magnetic response in Fig.~\ref{nambu-fig3}(b) is basically equivalent to unpolarized neutron scattering.
The chiral term $M_{\rm ch}$ is plotted in Fig.~\ref{nambu-fig3}(c).
The dispersion is the same as in the magnetic response, but the sign (color) of the signal distinguishes the polarization of the magnon modes.
Note that Fig.~\ref{nambu-fig3} summarizes the data obtained at 293~K, which shows softening behavior compared with Fig.~\ref{fig:7} obtained at 20~K as explained earlier.
The red acoustic mode has the ``positive'' polarization (counterclockwise with respect to the field), whereas the blue optical mode is the exchange-split mode that precesses in the opposite (clockwise) direction.

We compare the measurements with the polarized neutron partial differential cross section calculated using atomistic spin dynamics with quantum statistics~\cite{Barker2016b,Barker2019}.
The exchange constants are taken from Princep {\it et al}.~\cite{Princep2017a} and scaled by $S^2$ ($S=5/2$).
With the convolution of the approximated instrument resolution, Fig.~\ref{nambu-fig3}(d) shows excellent agreement with the experiments.
Calculation with the parameter set from Ref.~30 also gives good agreement, since both yield almost identical magnon dispersion relation below 35~meV.

\begin{figure*}[t]
	\begin{center}
		\includegraphics[width=0.8\linewidth,clip]{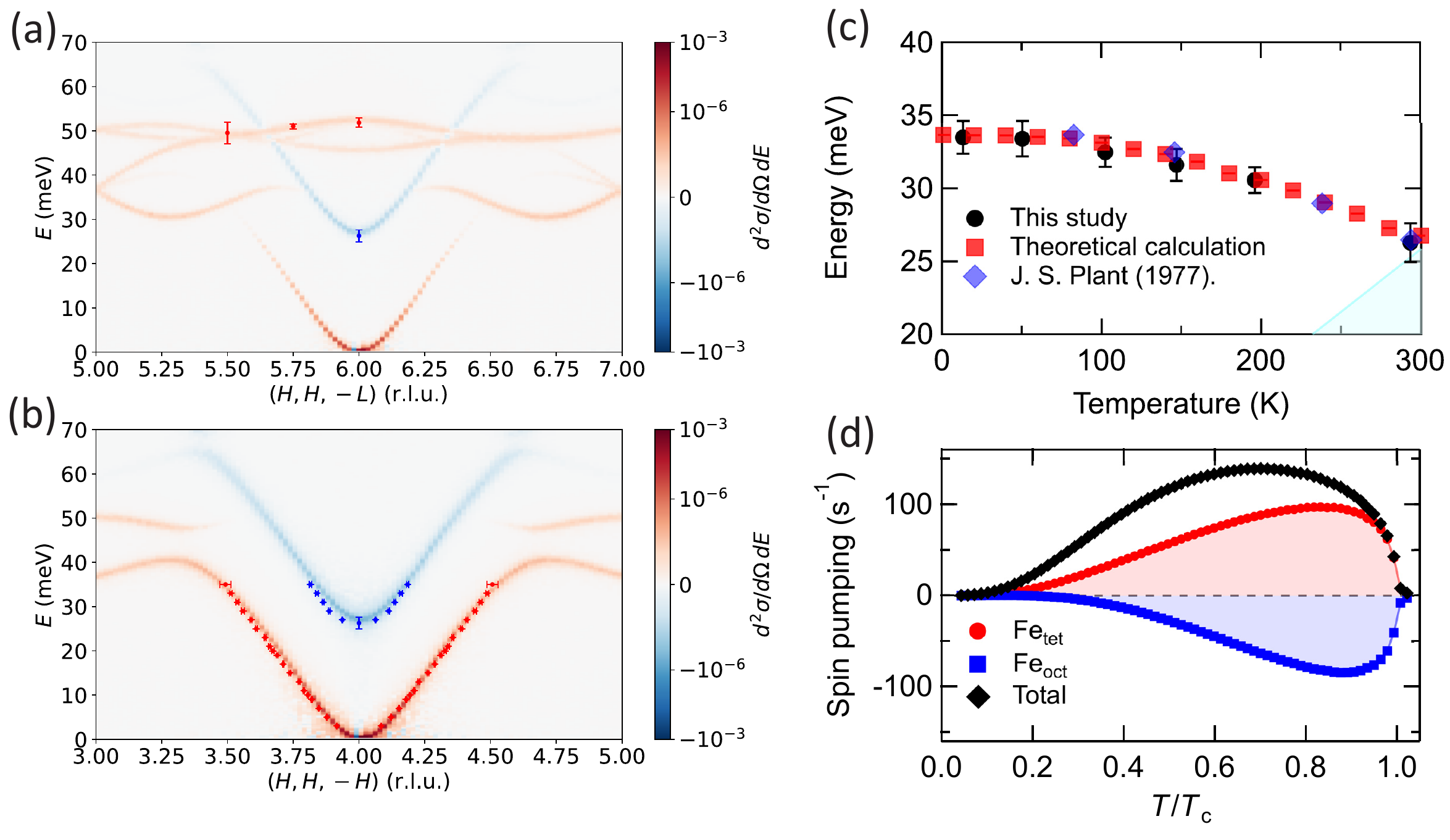}
	\end{center}
	\caption{(Color online) (a), (b) Calculated partial differential scattering cross sections overlaid with experimentally estimated peak positions at 293~K. $(H,H,-L)$ in (a) and $(H,H,-H)$ in (b) span the ranges $(5,5,-3)$ to $(7,7,-5)$ and $(3,3,-3)$ to $(5,5,-5)$, respectively. (c) Temperature dependence of the estimated optical gap value compared with the calculation and the previous results~\cite{Plant1977}. The shaded area marks $E\le k_{\rm B}T$. (d) Calculated $T/T_{\rm C}$ dependence of the thermal spin pumping from Y$_3$Fe$_5$O$_{12}$ by the Fe$_{\rm tet}$ and Fe$_{\rm oct}$ sites, and the total. Reprinted with permission from Nambu {\it et al}.~\cite{Nambu2020} ({\copyright} 2020 The American Physical Society).}
	\label{nambu-fig4} 
\end{figure*}

We measured a large number of points on the two magnon branches and also measured an optical mode with positive polarization by moving to the $(6,6,-4)$ Brillouin zone (Fig.~\ref{nambu-fig4}(a)).
Peaks were extracted using resolution-convoluted fits with the recently developed Eckold--Sobolev-type resolution function~\cite{Eckold2014}.
We found almost perfect agreement between the experiment and theory for both the nearly flat magnon mode at 50~meV around $(6,6,-4)$ (Fig.~\ref{nambu-fig4}(a)) and the acoustic and optical modes below 35~meV around $(4,4,-4)$ (Fig.~\ref{nambu-fig4}(b)).
The agreement from a low temperature to room temperature validates the low-temperature parameterization of the exchange coupling constants~\cite{Princep2017a}.
The magnon polarization of the localized (flat) mode is positive, in agreement with the calculations, highlighting the ability to measure polarization anywhere in the reciprocal space.

The optical gap is important for the thermodynamic and transport properties of YIG around and above room temperature, including the spin Seebeck effect.
Magnon modes are thermally occupied below $E=k_{\rm B}T$ (shaded area in Fig.~\ref{nambu-fig4}(c)).
At low temperatures, only the acoustic mode is occupied, but at room temperature and above, the optical mode with the opposite polarization becomes occupied. 
The strength of the spin Seebeck signal is proportional to the product of the integrated energy and the magnon polarization, to which the acoustic and optical modes contribute with opposite signs.
The observed maximum of the spin Seebeck voltage in YIG near room temperature~\cite{Kikkawa2015} has been interpreted in terms of this competition.
This is also theoretically illustrated in Fig.~\ref{nambu-fig4}(d), in which the total spin-pumping signal is clearly not the sum of these from the Fe$_{\rm tet}$ and Fe$_{\rm oct}$ moments, but the spin Seebeck voltage drops much faster than the magnetization with increasing temperature.
Whereas a theoretical treatment including interactions on the Pr/YIG interface is not yet available, the optical modes are thus expected to play an important role in the spin Seebeck voltage.
The optical modes might also explain the observation of reduced magnon conductivity~\cite{Wimmer2018}.

\section{Discussion}

\begin{table*}[t]
	\caption{Reported exchange parameters in the unit of meV for Y$_3$Fe$_5$O$_{12}$. $J_i^{jk}$ stands for the $i$th neighbor interaction between the $j$ and $k$ sites. Note that $J_3^{aa}$ and $J_3^{aa\prime}$ with identical displacement have separate superexchange pathways. The positive and negative signs respectively correspond to antiferromagnetic and ferromagnetic interactions.}
	\label{table-J}
	\begin{center}
		\begin{tabular}{lccccccc}
			\hline
			& $J_1^{ad}$ & $J_2^{dd}$ & $J_3^{aa}$ & $J_3^{aa\prime}$ & $J_4^{ad}$ & $J_5^{dd}$ & $J_6^{aa}$\\
			\hline
			Wojtowicz (1964)~\cite{Wojtowicz} & 5.56 & 0.56 & 0 & 0 & -- & -- & --\\
			Plant (1977)~\cite{Plant1977} & 6.86 & 1.38 & 1.38 & 1.38 & -- & -- & --\\
			Plant (1983)~\cite{Plant2} & 6.4 & 0.9 & 0 & 0 & 0.46 & 0.28 & 1.5\\
			Cherepanov {\it et al}. (1993)~\cite{Cherepanov1993} & 6.87 & 2.3 & 0.65 & 0.65 & -- & -- & --\\
			Princep {\it et al}. (2017)~\cite{Princep2017a} & 6.8(2) & 0.52(4) & 0.0(1) & 1.1(3) & -0.07(2) & 0.47(8) & -0.09(5)\\
			Xie {\it et al}. (2017)~\cite{Xie} & 4.774 & 0.308 & 0.144 & 0.144 & 0.326 & 0.358 & 0.008\\
			Shamoto {\it et al}. (2018)~\cite{Shamoto2018} & 5.80(14) & 0.70(16) & 0.0(1) & 0.0(1) & -- & -- & --\\
			\hline
		\end{tabular}
	\end{center}
\end{table*}

\begin{figure*}[h]
	\begin{center}
		\includegraphics[width=\linewidth,clip]{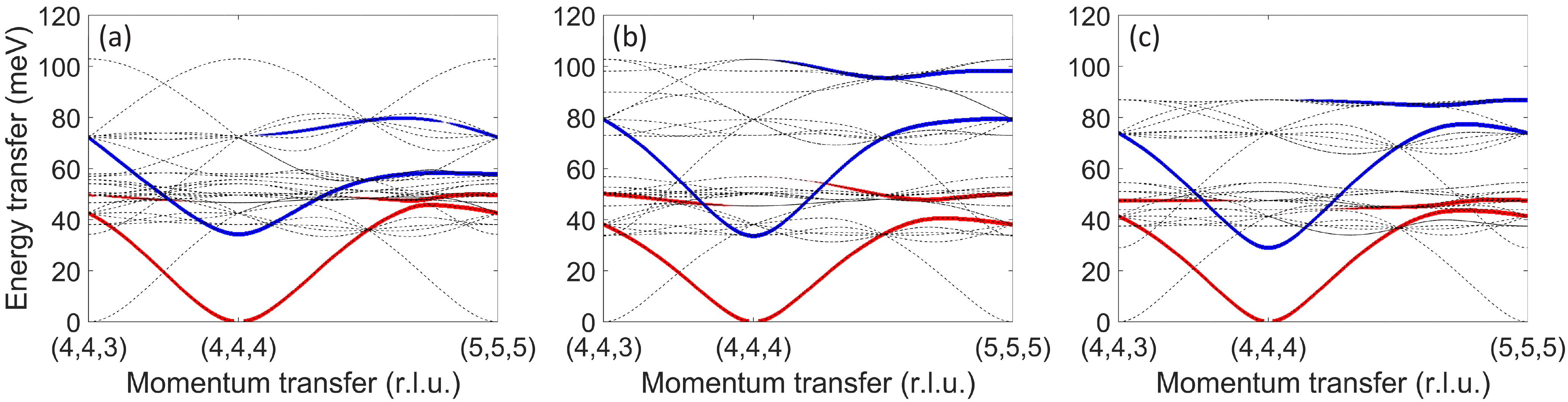}
	\end{center}
	\caption{(Color online) Spin-wave calculations for Y$_3$Fe$_5$O$_{12}$ based on the exchange parameters from (a) Plant (1983)~\cite{Plant2}, (b) Princep {\it et al}. (2017)~\cite{Princep2017a},  and (c) Shamoto {\it et al}. (2018)~\cite{Shamoto2018}. Dashed curves represent the dispersion relations, and solid curves with the color correspond to the sign and magnitude of the chiral correlation function, i.e., the magnon polarization.}
	\label{spinwavecalc} 
\end{figure*}

Both inelastic {\it unpolarized} and {\it polarized} neutron scattering measurements have been performed to study the magnetic excitation in YIG.
Each measurement has its strengths and weaknesses.
The magnon polarization for each magnon branch can only be observed by inelastic polarized neutron scattering.
However, the polarization of neutrons strongly reduces the count rate and limits the accessible energy range due to the performance of polarization devices.
Moreover, the energy resolution is sometimes sacrificed to maximize the neutron flux.
Unpolarized neutron scattering, on the other hand, is suitable for studying high-energy and ultralow-energy spin-wave spectra with relatively high resolution.
Another typical example is the determination of exchange parameters.

Fittings have been made on inelastic unpolarized neutron scattering results~\cite{Xie,Princep2017a,Shamoto2018}, and representative parameter sets are summarized in Table~\ref{table-J}.
We compared the calculated spin-wave spectra in Fig.~\ref{spinwavecalc} and found that they are consistent with each other for two major branches below 40~meV.
Differences are, however, visible for higher energy transfers around the maxima of the acoustic and optical modes.
The obtained parameters depend on the energy transfer regime used during the fitting.
The spin-wave spectrum should be measured in the whole energy range with a reasonably high $E$-resolution for precise fitting.
This has been a challenging condition for any inelastic neutron scattering, especially for inelastic polarized neutron scattering.
This is due to the limitation of polarizers for high-energy neutron beams.
The $^3$He spin filter and dynamic nuclear polarization methods have been developed for the neutron polarization in higher energy transfers.

In the inelastic unpolarized neutron scattering measurement, absolute-scale intensity estimation has played an important role in discoveries.
The ultralow-energy spin-wave dispersion with a Zeeman energy gap under an external magnetic field was determined from the energy dependence of the absolute intensity.
The magnon mode number of YIG was also determined to be unity at low energies by the intensity estimation.
However, it was impossible to determine the polarization mode by this method.
The clear demonstration of two different polarizations in the acoustic and optical modes by inelastic polarized neutron scattering measurement shows the importance of two-mode mixing at high temperatures.
Although the mixing plays an important role in the spin Seebeck effect, the present detailed study on the spin-wave by inelastic unpolarized neutron scattering reveals various anomalies depending on the magnetic field.
These anomalies were overlooked in the previous measurements.
Each technique was used to reveal novel features of YIG via its advantages.
All these techniques are applicable to other magnetic systems.

Through our first attempts to observe magnon polarization~\cite{Nambu2020}, the magnon polarization was found to be a fundamental property of matter of relevance to spintronic phenomena.
Our technique~\cite{Nambu2020} to resolve the magnon polarization is also applicable for other ferrimagnetic systems.
For instance, Gd$_3$Fe$_5$O$_{12}$ shows a sign change in the spin Seebeck voltage~\cite{Geprags2016}, in which modes with different polarizations are thought to exist close together.
A magnon polarization analysis of rare-earth iron garnets could also help to understand the observed magnon spin currents~\cite{Cramer2017}.
In a magnetically soft material such as YIG, the magnon polarization is nearly circular; however, YIG is very amenable to doping, and magnetic anisotropies can also be introduced.
Strong anisotropies, as well as local anisotropy in the tetrahedral and octahedral sites, may couple magnons with opposite polarization, thereby causing ellipticity and anticrossings between optical and acoustic modes.
This ``magnon squeezing''~\cite{Kamra2016} may be essential for applications of magnets in quantum information and can be measured by this technique.

\section{Conclusions}

We have studied the basic characteristics of the quintessential magnet YIG using neutrons.
Although YIG has been believed to have the space group $Ia\bar{3}d$, our detailed crystal structure refinement showed distortion to the trigonal $R\bar{3}$~\cite{Shamoto2018}.
Unpolarized neutron scattering experiments revealed magnetic excitations~\cite{Shamoto2018,Shamoto2020} ranging from high (100~meV) to ultralow energy (10~$\mu$eV).
Through linearized spin-wave analysis, the nearest-neighbor exchange interactions, $J_{aa}$, $J_{ad}$, and $J_{dd}$, were estimated, and the stiffness constant $D$ was found to be consistent within this approximation.

We also measured the polarization of magnons, which is an important degree of freedom in magnets yet hitherto untested, in collinear ferrimagnetic YIG and found quantitative agreement with the theory~\cite{Nambu2020}.
The magnon polarization can easily be squeezed by spin anisotropy and/or any mixing between the acoustic and optical modes.
Our first attempt to measure the pure polarization using YIG can therefore be a textbook case for magnon polarization observation.
We anticipate that valuable information can be gained from similar measurements on other ferrimagnets.
Theories discussing the role of magnon polarization in spintronics are now appearing~\cite{Kamra2017}, and our direct measurement of the magnon polarization has thus demonstrated the importance of neutron scattering for the next generation of spintronics and magnonics.

\begin{acknowledgments}
We acknowledge the following individuals for fruitful discussions: M. Akatsu, J. Barker, S.~E. Barnes, G.~E.~W. Bauer, L.-J. Chang, M. Enderle, H. Endo, M. Fujita, J. Ieda, Y. Inamura, T.~U. Ito, R. Kajimoto, K. Kakurai, T. Kikkawa, Y. Kobayashi, K. Kodama, M. Kofu, C.-H. Lee, S. Maekawa, M. Matsuura, M. Mori, T. Moyoshi, K. Munakata, M. Nakamura, A. Nakao, Y. Nemoto, T. Oda, T. Ohhara, S. Ohira-Kawamura, H. Onishi, Y. Ohnuma, E. Saitoh, N. Sato, K. Shibata, Y. Shiomi, S. Toth, J.~M. Tranquada, T. Weber, B. Winn, H. Yamauchi, Y. Yasui, and T. Ziman.
We also thank the CROSS sample environment team and M.~B\"ohm for their experimental assistance, and M. Usami and Y. Baba in the JAEA technical support team.

The work at J-PARC was performed under proposals 2012B0134, 2015A0174 (BL01), 2014B0157, 2015I0002, 2016A0318, 2017L0301 (BL02), and 2013B0278 (BL14).
The work at ILL was performed under project 4-01-1559 (doi:10.5291/ILL-DATA.4-01-1559).
This work was supported by JSPS (Nos.~JP21H03732, JP25287094, JP16K05424, JP16H04007, JP17H05473, JP19H04683, JP17H06137), JST (No.~JPMJFR202V) Iketani Science and Technology Foundation, and the Graduate Program in Spintronics at Tohoku University.
\end{acknowledgments}

\profile{Yusuke Nambu}{obtained his Ph.D. degree from Kyoto University (2009). He was a JSPS research fellow (2006--2009), a JSPS postdoctoral fellow for research abroad at Johns Hopkins University (2009--2010), and a guest researcher at NIST Center for Neutron Research (2009--2010). After serving as an assistant professor at the University of Tokyo (2010--2012) and Tohoku University (2012--2015), he has been an associate professor since 2015, a Tohoku University Distinguished Researcher since 2020, and a JST FOREST researcher since 2021. His research is currently focused on magnetism in spintronics, frustrated magnetism, and iron-based superconductivity using neutron diffraction/scattering techniques.}

\profile{Shin-ichi Shamoto}{was born in Nagoya in 1960. He received B.S. (1983) and M.S. (1985) degrees from Kyoto University, and a D.S. (1990) degree from the University of Tokyo. He was a research associate (1991--1995) and a lecturer (1995--1996) at Nagoya University, an associate professor (1996--2004) at Tohoku University, and a principal researcher (2004--2008) and a senior principal researcher (2008--2020) at the Japan Atomic Energy Agency. He is currently a science coordinator (2020--) at CROSS and a visiting chair professor (2020--) at National Cheng Kung Univeristy in Taiwan. His research interests include neutron scattering on nanostructures and dynamics of functional materials such as iron-based superconductors, nanomaterials, and magnetic materials.}


\begin{thebibliography}{99}
\bibitem{Silsbee1979}
R.~H.~Silsbee, A. Janossy, and P. Monod, Phys. Rev. B {\bf 19}, 4382 (1979).
\bibitem{Mizukami2002}
S. Mizukami, Y. Ando, and T. Miyazaki, Phys. Rev. B {\bf 66}, 104413 (2002).
\bibitem{Tserkovnyak2002}
Y. Tserkovnyak, A. Brataas, and G.~E.~W. Bauer, Phys. Rev. Lett. {\bf 88}, 117601 (2002).
\bibitem{Kato2004}
Y.~K. Kato, R.~C. Myers, A.~C. Gossard, and D.~D. Awschalom, Science {\bf 306}, 1910 (2004).
\bibitem{Wunderlich2005}
J. Wunderlich, B. Kaestner, J. Sinova, and T. Jungwirth, Phys. Rev. Lett. {\bf 94}, 047204 (2005).
\bibitem{Prins1995}
M.~W.~J. Prins, H. van Kempen, H. van Leuken, R.~A. de Groot, W.~V. Roy, and J.~D. Boeck, J. Phys: Condens. Matter {\bf 7}, 9447 (1995).
\bibitem{Uchida2008}
K. Uchida, S. Takahashi, K. Harii, J. Ieda, W. Koshibae, K. Ando, S. Maekawa, and E. Saitoh, Nature {\bf 455}, 778 (2008).
\bibitem{Slachter2010}
A. Slachter, F.~L. Bakker, J.-P. Adam, and B.~J. van Wees, Nat. Phys. {\bf 6}, 879 (2010).
\bibitem{Cornelissen2015}
L.~J. Cornelissen, J. Liu, R.~A. Duine, J. Ben Youssef, and B.~J. van Wees, Nat. Phys. {\bf 11}, 1022 (2015).
\bibitem{Azevedo2005}
A. Azevedo, O. Alves Santos, G.~A. Fonseca Guerra, R.~O. Cunha, R. Rodr\'iguez-Su\'arez, and S.~M. Rezende, J. Appl. Phys. {\bf 97}, 10C715 (2005).
\bibitem{Saitoh2006}
E. Saitoh, M. Ueda, H. Miyajima, and G. Tatara, Appl. Phys. Lett. {\bf 88}, 182509 (2006).
\bibitem{Valenzuela2006}
S.~O. Valenzuela and M. Tinkham, Nature {\bf 442}, 176 (2006).
\bibitem{Kimura2007}
T. Kimura, Y. Otani, T. Sato, S. Takahashi, and S. Maekawa, Phys. Rev. Lett. {\bf 98}, 156601 (2007).
\bibitem{Wu2013}
M. Wu and A. Hoffmann, Solid State Phys. {\bf 64}, 1 (2013).
\bibitem{Tabuchi2015}
Y. Tabuchi, S. Ishino, A. Noguchi, T. Ishikawa, R. Yamazaki, K. Usami, and Y. Nakamura, Science {\bf 349}, 405 (2015).
\bibitem{Chang2014}
H. Chang, P. Li, W. Zhang, T. Liu, A. Hoffmann, L. Deng, and M. Wu, IEEE Magn. Lett. {\bf 5}, 6700104 (2014).
\bibitem{Shamoto2018}
S. Shamoto, T. U. Ito, H. Onishi, H. Yamauchi, Y. Inamura, M. Matsuura, M. Akatsu, K. Kodama, A. Nakao, T. Moyoshi, K. Munakata, T. Ohhara, M. Nakamura, S. Ohira-Kawamura, Y. Nemoto, and K. Shibata, Phys. Rev. B {\bf 97}, 054429 (2018).
\bibitem{Shamoto2020}
S. Shamoto, Y. Yasui, M. Matsuura, M. Akatsu, Y. Kobayashi, Y. Nemoto, and J. Ieda, Phys. Rev. Res. {\bf 2}, 033235 (2020).
\bibitem{Nambu2020}
Y. Nambu, J. Barker, Y. Okino, T. Kikkawa, Y. Shiomi, M. Enderle, T. Weber, B. Winn, M. Graves-Brook, J.~M. Tranquada, T. Ziman, M. Fujita, G.~E.~W. Bauer, E. Saitoh, and K. Kakurai, Phys. Rev. Lett. {\bf 125}, 027201 (2020).
\bibitem{Kikkawa2015}
T. Kikkawa, K. Uchida, S. Daimon, Z. Qiu, Y. Shiomi, and E. Saitoh, Phys. Rev. B {\bf 92}, 064413 (2015).
\bibitem{Kikkawa2016}
T. Kikkawa, K. Shen, B. Flebus, R. A. Duine, K. Uchida, Z. Qiu, G.~E.~W. Bauer, and E. Saitoh, Phys. Rev. Lett. {\bf 117}, 207203 (2016).
\bibitem{Cherepanov}
V. Cherepanov, I. Kolokolov, and V. L'vov, Phys. Rep.  {\bf 229}, 81 (1993).
\bibitem{Rodic}
D. Rodic, M. Mitric, R. Tellgren, H. Rundlof, and A. Kremenovic, J. Magn. Magn. Mater. {\bf 191}, 137 (1999).
\bibitem{Kimura} 
S. Kimura and I. Shindo, J. Cryst. Growth  {\bf 41}, 192 (1977).
\bibitem{SENJU}
T. Ohhara, R. Kiyanagi, K. Oikawa, K. Kaneko, T. Kawasaki, I. Tamura, A. Nakao, T. Hanashima, K. Munakata, T. Moyoshi, T. Kuroda, H. Kimura, T. Sakakura, C.-H. Lee, M. Takahashi, K. Ohshima, T. Kiyotani, Y. Noda, and M. Arai, J. Appl. Crystallogr. {\bf49}, 120 (2016). 
\bibitem{FullProf}
J. Rodriguez-Carvajal, Physica B (Amsterdam, Neth.) {\bf192}, 55 (1993).
\bibitem{STARGazer}
T. Ohhara, K. Kusaka, T. Hosoya, K. Kurihara, K. Tomoyori, N. Niimura, I. Tanaka, J. Suzuki, T. Nakatani, T. Otomo, S. Matsuoka, K. Tomita, Y. Nishimaki, T. Ajima, and S. Ryufuku, Nucl. Instrum. Methods Phys. Res., Sect. A {\bf600}, 195 (2009).
\bibitem{VESTA}
K. Momma and F. Izumi, J. Appl. Crystallogr. {\bf 41}, 653 (2008).
\bibitem{Plant1977}
J.~S. Plant, J. Phys. C: Solid State Phys. {\bf 10}, 4805 (1977).
\bibitem{Plant2}
J.~S. Plant, J. Phys. C: Solid State Phys. {\bf 16}, 7037 (1983).
\bibitem{Serga} 
A.~A. Serga, A.~V. Chumak, and B. Hillebrands, J. Phys. D: Appl. Phys. {\bf 43}, 264002 (2010).
\bibitem{Princep2017a}
A.~J. Princep, R.~A. Ewings, S. Ward, S. T\'oth, C. Dubs, D. Prabhakaran, and A.~T. Boothroyd, npj Quantum Mater. {\bf 2}, 63 (2017).
\bibitem{Barker}
J. Barker and G.~E.~W. Bauer, Phys. Rev. Lett.  {\bf 117}, 217201 (2016).
\bibitem{spinw}
S. Toth and B. Lake, J. Phys.: Condens. Matter  {\bf27}, 166002 (2015). 
\bibitem{Horace}
R. A. Ewings, A. Buts, M. D. Le, J. van Duijn, I. Bustinduy, and T. G. Perring, Nucl. Instrum. Methods Phys. Res., Sect.
A {\bf834}, 132 (2016).
\bibitem{Kajimoto} 
R. Kajimoto, M. Nakamura, Y. Inamura, F. Mizuno, K. Nakajima, S. Ohira-Kawamura, T. Yokoo, T. Nakatani, R. Maruyama, K. Soyama, K. Shibata, K. Suzuya, S. Sato, K. Aizawa, M. Arai, S. Wakimoto, M. Ishikado, S. Shamoto, M. Fujita, H. Hiraka, K. Ohoyama, K. Yamada, and C.-H. Lee, J. Phys. Soc. Jpn. {\bf 80}, SB025 (2011).	
\bibitem{Nakajima}
K. Nakajima, S. Ohira-Kawamura, T. Kikuchi, M. Nakamura, R. Kajimoto, Y. Inamura, N. Takahashi, K. Aizawa, K. Suzuya, K. Shibata, T. Nakatani, K. Soyama, R. Maruyama, H. Tanaka, W. Kambara, T. Iwahashi, Y. Itoh, T. Osakabe, S. Wakimoto, K. Kakurai, F. Maekawa, M. Harada, K. Oikawa, R. E. Lechner, F. Mezei, and M. Arai, J. Phys. Soc. Jpn. {\bf 80}, SB028 (2011).
\bibitem{Nakamura} 
M. Nakamura, R. Kajimoto, Y. Inamura, F. Mizuno, M. Fujita, T. Yokoo, and M. Arai, J. Phys. Soc. Jpn. {\bf 78}, 093002 (2009).
\bibitem{Shibata}  
K. Shibata, N. Takahashi, Y. Kawakita, M. Matsuura, T. Yamada, T. Tominaga, W. Kambara, M. Kobayashi, Y. Inamura, T. Nakatani, K. Nakajima, and M. Arai, JPS Conf. Proc. {\bf 8}, 036022 (2015).
\bibitem{Utsusemi}
Y. Inamura, T. Nakatani, J. Suzuki, and T. Otomo, J. Phys. Soc. Jpn. {\bf 82}, SA031 (2013).
\bibitem{Wojtowicz}
P. J. Wojtowicz, Phys. Lett.  {\bf11}, 18 (1964). 
\bibitem{Geller}
S. Geller, H.~J. Williams, G.~P. Espinosa, and R.~C. Sherwood, Bell Syst. Tech. J. {\bf43}, 565 (1964).
\bibitem{Xie} 
L.-S. Xie, G.-X. Jin, L. He, G. E. W. Bauer, J. Barker, and K. Xia, Phys. Rev. B {\bf 95}, 014423 (2017).
\bibitem{Man2017} 
H. Man, Z. Shi, G. Xu, Y. Xu, X. Chen, S. Sullivan, J. Zhou, K. Xia, J. Shi, and P. Dai, Phys. Rev. B {\bf 96}, 100406(R) (2017).
\bibitem{Cherepanov1993}
V. Cherepanov, I. Kolokolov, and V. L'vov, Phys. Rep. {\bf 229}, 81 (1993).
\bibitem{Srivastava}
C. M. Srivastava and R. Aiyar, J. Phys. C {\bf 20}, 1119 (1987).
\bibitem{Shirane}
G. Shirane, S. M. Shapiro, and J. M. Tranquada, {\it Neutron Scattering with a Triple-Axis Spectrometer} (Cambridge Univ. Press, 2002).
\bibitem{Park}
J. Jeong, M. D. Le, P. Bourges, S. Petit, S. Furukawa, S.-A. Kim, S. Lee, S.-W. Cheong, and J.-G. Park, Phys. Rev. Lett. {\bf 113}, 107202 (2014).
\bibitem{LeCraw}
R.~C. LeCraw and L.~R. Walker, J. Appl. Phys. {\bf 32}, S167 (1961).
\bibitem{Kittel}
C. Kittel, {\it Quantum Theory of Solids} (John Wiley \& Sons, Inc., New York, 1987).
\bibitem{Dillon}
J.~F. Dillon Jr., Phys. Rev. {\bf 105}, 759 (1957).
\bibitem{Yamasaki2009}
Y. Yamasaki, Y. Kohara, and Y. Tokura, Phys. Rev. B {\bf 80}, 140412(R) (2009).
\bibitem{Kohara}
Y. Kohara, Y. Yamasaki, Y. Onose, and Y. Tokura, Phys. Rev. B {\bf 82}, 104419 (2010).
\bibitem{Xiao2010}
J. Xiao, G.~E.~W. Bauer, K. Uchida, E. Saitoh, and S. Maekawa, Phys. Rev. B {\bf 81}, 214418 (2010).
\bibitem{Schlomann1960}
E. Schl\"omann, in Solid State Physics in Electronics and Telecommunication (Academic Press, New York, Vol. 3, p. 322, 1960).
\bibitem{Chatterji2006}
T. Chatterji, Neutron Scattering from Magnetic Materials (Elsevier, Amsterdam, 2006).
\bibitem{Moon1969}
R.~M. Moon, T. Riste, and W.~C. Koehler, Phys. Rev. {\bf 181}, 920 (1969).
\bibitem{Kakurai1984}
K. Kakurai, R. Pynn, B. Dorner, and M. Steiner, J. Phys. C: Solid State Phys. {\bf 17}, L123 (1984).
\bibitem{Maleyev1995}
S.~V. Maleyev, Phys. Rev. Lett. {\bf 75}, 4682 (1995).
\bibitem{Loire2011}
M. Loire, V. Simonet, S. Petit, K. Marty, P. Bordet, P. Lejay, J. Ollivier, M. Enderle, P. Steffens, E. Ressouche, A. Zorko, and R. Ballou, Phys. Rev. Lett. {\bf 106}, 207201 (2011).
\bibitem{Roessli2002}
B. Roessli, P. B\"oni, W.~E. Fischer, and Y. Endoh, Phys. Rev. Lett. {\bf 88}, 237204 (2002).
\bibitem{Lorenzo2007a}
J.~E. Lorenzo, C. Boullier, L.~P. Regnault, U. Ammerahl, and A. Revcolevschi, Phys. Rev. B {\bf 75}, 054418 (2007).
\bibitem{Blume1963}
M. Blume, Phys. Rev. {\bf 130}, 1670 (1963).
\bibitem{Maleyev1963}
S.~V. Maleyev, V.~G. Baryakhtar, and A. Suris, Sov. Phys. Solid State {\bf 4}, 2533 (1963).
\bibitem{Nambu-arXiv}
Y. Nambu, M. Enderle, T. Weber, and K. Kakurai, arXiv:2006.15758.
\bibitem{IN20}
J. Kulda, J. \v{S}aroun, P. Courtois, M. Enderle, M. Thomas, and P. Flores, Appl. Phys. A {\bf 74}, S246 (2002).
\bibitem{Barker2016b}
J. Barker and G.~E.~W. Bauer, Phys. Rev. Lett. {\bf 117}, 217201 (2016).
\bibitem{Barker2019}
J. Barker and G.~E.~W. Bauer, Phys. Rev. B {\bf 100}, 140401(R) (2019).
\bibitem{Eckold2014}
G. Eckold and O. Sobolev, Nucl. Instrum. Methods Phys. Res. A {\bf 752}, 54 (2014).
\bibitem{Wimmer2018}
T. Wimmer, M. Althammer, L. Liensberger, N. Vlietstra, S. Gepr{\"a}gs, M. Weiler, R. Gross, and H. Huebl, arXiv:1812.01334.
\bibitem{Geprags2016}
S. Gepr\"ags, A. Kehlberger, F. Della Coletta, Z. Qiu, E.~J. Guo, T. Schulz, C. Mix, S. Meyer, A. Kamra, M. Althammer, H. Huebl, G. Jakob, Y. Ohnuma, H. Adachi, J. Barker, S. Maekawa, G.~E.~W. Bauer, E. Saitoh, R. Gross, S.~T.~B. Goennenwein, and M. Kl\"aui, Nat. Commun. {\bf 7}, 10452 (2016).
\bibitem{Cramer2017}
J. Cramer, E. Guo, S. Gepr{\"a}gs, A. Kehlberger, Y.~P. Ivanov, K. Ganzhorn, F. Della Coletta, M. Althammer, H. Huebl, R. Gross, J. Kosel, M. Kl{\"a}ui, S.~T.~B. Goennenwein, Nano Lett. {\bf 17}, 3334 (2017).
\bibitem{Kamra2016}
A. Kamra and W. Belzig, Phys. Rev. Lett. {\bf 116}, 146601 (2016).
\bibitem{Kamra2017}
A. Kamra, U. Agrawal, and W. Belzig, Phys. Rev. B {\bf 96}, 020411(R) (2017).
\end{thebibliography}
\end{document}